# Beyond Wellbeing Apps: Co-Designing Immersive, Embodied, and Collective Digital Wellbeing Interventions for Healthcare Professionals

Zheyuan Zhang*
Dyson School of Design Engineering
Imperial College London
London, United Kingdom
zheyuan.zhang17@imperial.ac.uk

Jingjing Sun
Dyson School of Design Engineering
Imperial College London
London, United Kingdom
j.sun23@imperial.ac.uk

Dorian Peters
Dyson School of Design Engineering
Imperial College London
London, United Kingdom
d.peters@imperial.ac.uk

Rafael A Calvo
Dyson School of Design Engineering
Imperial College London
London, United Kingdom
r.calvo@ic.ac.uk

## Abstract

Healthcare professionals (HCPs) face increasing levels of stress and burnout. Technological wellbeing interventions provide accessible and flexible support for HCPs. While most studies have focused on mobile- and web-based programs, alternative technologies like virtual reality (VR), augmented reality (AR), tangible interfaces, and embodied technologies are emerging as engaging and effective tools for wellbeing interventions. However, there is still a lack of research on how such technologies are perceived among HCPs. This study explored HCPs' perceptions and preferences for various types of wellbeing technologies, by conducting a 2-phase co-design study involving 26 HCPs in idea generation, concept evaluation, prototype testing, and design iteration. From our findings, HCPs highly valued the potential of technologies to support mental health with immersive, embodied, and collective experiences. Furthermore, we provided design recommendations for wellbeing technologies for HCPs that sustain user engagement by meeting their needs for autonomy, competence, and relatedness in the experiences.

*Corresponding author.

## 1 Introduction

Healthcare professionals (HCPs) are consistently taxed by increasing job demands and a complex working environment, which makes them vulnerable to burnout and poor mental wellbeing [30, 40]. COVID-19 and its residual impacts on healthcare systems across the world have further burdened HCPs and put their mental health at risk [70]. Additionally, burnout and poor mental health among HCPs can profoundly impact the quality of care they deliver, resulting in more medical errors and patient dissatisfaction [35].

Numerous non-technological interventions have been shown to mitigate staff burnout and promote wellbeing in HCPs. These include mindfulness-based stress reduction sessions [20, 65], peer support programs [30], psychosocial skills training [86] and physical relaxation activities [27, 75]. However, the impact of these interventions is limited by constraints on scaling up sessions that require facilitation and a set schedule. Technology-facilitated wellbeing interventions provide greater access and have attracted growing research interest [77]. Among the research on digitally delivered well-being interventions (including those specifically developed for HCPs), the majority have focused on programs delivered via mobile devices and computers [22, 111]. Although interventions delivered through text messaging, apps and websites offer private and flexible support [36, 78], a lack of user engagement and high attrition have been reported among targeted users, including HCPs [67]. Research has identified some common issues, including a general focus on content that fails to engage, limited content relevance to users, a lack of human support and face-to-face connections, and difficulty integrating these tools into daily routines [16, 60, 67].

In parallel, there is growing recognition of the potential value of interventions facilitated by other technologies beyond apps and websites [5]. Virtual reality (VR) [44, 81], augmented reality (AR) [123], embodied and tangible technologies [33, 39, 43, 115], conversational agents [73, 104], and ambient technologies [128] have all demonstrated promising outcomes in Human-Computer Interaction (HCI) and mental health research. These alternative technologies offer immersive, embodied and engaging approaches to stress management and mental wellbeing [3, 33, 71]. Indeed, VR-based interventions, for example, have the potential to effectively reduce workplace stress and anxiety in healthcare settings [2, 44, 86] and increase HCPs' engagement compared with smartphone apps [86]. Furthermore, wellbeing interventions using alternative forms of technology often provide on-site embodiment and allow easier integration with daily work routines [114]. In this paper, we use the term 'alternative technologies' to collectively refer to commercially available technologies beyond common mobile and computing devices, including VR, AR, tangible interfaces, and embodied systems. These technologies are distinct in that they afford immersion and/or embodiment through alternative viewing and interaction modalities, enabling more engaging and distinct experiences compared to traditional digital mental health tools [5, 8].

However, despite the promise of alternative technologies for mental wellbeing, there remains a considerable gap in the literature to show whether they might be realistically incorporated into healthcare settings. Little is known about how such alternative wellbeing technologies would be perceived by HCPs, or what designers and HCI researchers should take into account when designing and delivering such interventions for HCPs.

We set out to explore the broad horizon of alternative wellbeing technologies for HCPs. Specifically, we aimed to address the following two research questions: 1) What are the preferences and expressed values of HCPs regarding the design of well-being technologies to support their mental health? 2) What design implications can be drawn from HCPs' perspectives to enhance their engagement with alternative forms of digital mental well-being support? To answer these questions, we worked with 26 HCPs in a tertiary hospital in China across a 2-phase co-design study. Phase 1 consisted of HCPs individually generating ideas for technology-facilitated mental wellbeing interventions. Participants demonstrated a strong desire for embodied, immersive, and collective experiences facilitated by technologies. After the first phase of workshops, we developed a medium-fidelity prototype of an embodied digital intervention that addressed the needs and preferences proposed in Phase 1. For Phase 2, the prototype was installed and used both for proof-of-concept evaluation and further idea-elicitation, allowing participants to experience the technology in their work environment and provide further feedback on the prototype, and ideate on how to design wellbeing technologies to sustain user engagement in the long-term.

The contributions of this paper are three-fold: 1) Through co-design with HCPs, we gain user insights on how alternative technologies would benefit wellbeing interventions; 2) We provide a detailed analysis of HCP perspectives and preferences for wellbeing technologies in the workplace; 3) We demonstrate the practical application of these insights through the development and implementation of evidence-based design prototypes.

## 2 Related Work
### 2.1 Wellbeing Interventions for Healthcare Professionals

Over the past decades, research has been focused on interventions to mitigate burnout and promote wellbeing among HCPs through both person-directed and organization-directed approaches [29, 122]. Organization-directed interventions often promote changes that could positively affect the organizational climate and create a better working experience, like the reduction of working hours and shifts, changes in management style, and the establishment of peer-support groups, staff educational programs, or modifications to clinical processes [11, 34, 49, 52]. On the other hand, person-directed interventions often aim to promote skills, knowledge and capabilities to cope with work distress and improve personal wellbeing [84]. Some broadly explored person-directed interventions include psychotherapy (often using cognitive-behavioral therapy (CBT) [4] and acceptance and commitment therapy (ACT)) [4, 9], stress management training [107], communication skills training [11], debrief sessions [50], and mindfulness-based programs [20, 65].

Among the evidence-based personal interventions, many incorporate face-to-face training and activities that take place within the organization. For example, stress management sessions like mind-body exercises held regularly within the workplace were found effective in helping HCPs reduce physical tension, psychological stress, and burnout [27, 42]. Similarly, mindfulness-based interventions and resilience training for HCPs often involve multiple weekly sessions organized by and held within hospitals [68, 84].

### 2.2 Digital Interventions for Burnout

Although traditional burnout and work wellbeing interventions are found to be effective among HCPs, limited access and flexibility are often cited as constraints [1]. As an alternative, digital interventions are more flexible and accessible and have been developed and tested [57, 125]. For example, Profit et al. tested a web-based resilience-enhancing program with a Randomized Controlled Trial (RCT) and demonstrated its efficacy in reducing HCP burnout [95]. Hersch et al. also conducted an RCT and evaluated a web-based system called *BREATHE*, which provided strategies and tools, including stress management, positive coping, and mental health self-assessment for nurses to effectively cope with burnout [56].

It's worth noting that while traditional burnout interventions often involve on-site group activities in the workplace, digital interventions developed for HCPs mostly focus on individual self-help using personal technologies like smartphones and computers. In their systematic review, Lopez-Del-Hoyo et al. found 27 studies on digital mental health support for HCPs, with a vast majority focused on mental health apps, text messages, and websites [67]. In another review, Adam et al. identified 7 studies on digital interventions for HCPs in the period of 2017 −2022, with 6 focused on interventions delivered through mobile apps and websites [1].

Focusing on self-help using personal technologies offers flexibility and easy access, while, as some studies argued, at the cost of user engagement [16, 57, 66]. Lack of engagement and high user attrition have been commonly reported in app- and website-based interventions. In a pilot study on a web-based mindfulness program for General Practitioners (GPs), for example, only 9.4% of participants completed at least one online mindfulness session [78]. Furthermore, user engagement in real-life settings can be lower than in clinical studies, which often incentivizes engagement [47]. To illustrate this, a review of 59 mental health apps on the market reported a median uptake rate of 4.0% and a 15-day retention rate of only 3.9% [12]. Jardine et al. investigated the reasons behind existing and potential users' low motivation to adopt and sustain engagement with digital mental health products [60]. They found that a lack of human connection, forgetfulness, and a lack of content relevancy were major engagement hindrances [60]. Other barriers identified by other studies include lack of time due to high demand for work, lack of usability, lack of social connectedness, and low level of integration into life [16, 31].

### 2.3 Alternative Technologies to Facilitate Wellbeing

Research has also focused on wellbeing interventions using other forms of technologies beyond apps and websites. Studies have explored mindfulness and relaxation assisted by VR, smart speakers, large displays, and tangible interfaces. For example, Arpaia et al. reviewed over 50 recent studies on mindfulness in VR, identifying its effects on pain, stress, and anxiety [3]. They concluded that VR is the most effective technology for enhancing user engagement and retention among all technology-facilitated mindfulness practices [3]. The additional use of wearable devices to provide biofeedback during sessions has also been found to positively impact mindfulness outcomes [32]. Other studies have explored using soundscapes [28], smart speakers [104], high-definition displays [127], embodied and tangible tools [71], and haptic feedback [120] to promote mindfulness and relaxation practices. A generally positive impact on user engagement and wellbeing outcomes was reported.

Furthermore, HCI and digital health researchers have explored physical exercise and mind-body practices facilitated by VR, motion sensing, and wearable devices, showing promising outcomes in mental wellbeing. Facilitating physical exercise using VR and embodied technologies, such as virtual hiking, cycling, and motion-tracking exergames, has been validated to reduce stress, support mental wellbeing, and promote positive behavioral change across multiple studies [13, 51, 59, 116]. Meanwhile, mind-body practices like yoga, Tai-chi, and Qigong – validated as complementary medicines to reduce stress, anxiety, and burnout [63] – have also been integrated with various technologies. For example, researchers have evaluated the feasibility, user acceptability, and potential benefits of mind-body practices assisted by VR [53], AR [53, 62], motion sensing [83, 121], tangible interface [80], and wearables [117].

Although evidence has been garnered on using these alternative technologies for wellbeing, their application to HCPs is still scant. Only a few studies have explored mental health interventions for HCPs using alternative technologies [45, 86]. For instance, Pascual et al. [86] compared VR- and app-based mindfulness interventions with 32 HCPs in an emergency department and discovered higher engagement rates and larger heart-rate variability improvement (indicating better stress reduction) for the VR intervention. Ferrer Costa et al. [45] developed an 8-week VR educational program for HCPs, and their pre-post pilot test found a significant reduction in burnout and high adherence rates among HCPs. Despite the initial and promising evidence, it remains unclear if and how new forms of wellbeing technologies can be applied in healthcare settings, and what the key design and implementation considerations for such interventions for HCPs should be.

## 3 Methods

Co-designing digital health and mental wellbeing interventions with end-users is strongly recommended by HCI researchers [82], as this approach effectively engages end-users and yields in-depth design insights [10, 72, 105]. To further investigate the integration of wellbeing technologies in healthcare settings and address our research questions, we conducted a two-phase co-design study with our target user group, HCPs. As explained in Figure 1, the first phase gathered user perspectives and design ideas for implementing alternative technologies to address burnout and promote wellbeing. The second phase focused on prototype testing and design refinement.

### 3.1 Procedures

*3.1.1 Phase 1: Concept generation workshops.* As shown in Figure 2, Phase 1 workshops were divided into three parts: a brief introduction on the aim and process of the workshops, mental health and technology concept familiarization, and idea generation and evaluation (see Appendix A for details of the workshop plan). After introductions, participants were asked to reflect briefly on the current ways they use to relax, and on wellbeing interventions they have had experience with. They were then briefly introduced to some of the evidence-based wellbeing and burnout interventions for HCPs. Afterward, the research team provided introductions with exemplary references (see Table 1) to familiarize the participants with a broad set of potential technologies that can be used to assist with mental health and wellbeing interventions. The references were provided to address the varying levels of digital and mental health literacy among HCPs [54].

Participants were encouraged to use the information provided as a reference while freely brainstorming any forms of intervention and technology. Finally, participants were provided with a design template with a 4-step structure: interventions, technology, scenario, and expectations (see Figure 3a). Participants were encouraged to utilize this structure as an optional support for ideation and to freely express their ideas in any way they preferred. At the end of the workshops, participants presented their ideas and were asked to rate their favorite ideas, with 3 votes allocated to each participant (see Figure 4a).

*3.1.2 Phase 2: Prototype testing and design iteration workshops.* Based on HCPs' feedback, their preference votes, and the results of the data analysis, one concept was selected to be developed into a medium-fidelity prototype. The second phase of workshops aimed to evaluate the HCPs' experience with the initial prototype and acquire in-depth feedback on their needs and preferences to inform future design improvements. The prototype was brought to the hospital and installed by the research team prior to the testing workshops. The workshops comprised two parts. Participants were initially invited to test and experience the prototype with their peers. They were then asked to provide detailed reflections on their experience and generate improved design iterations that help the intervention to engage HCPs in the long term (see Table 4 (Appendix A) for details). To assist participants in the design iteration, the research team provided a set of 3 design templates that could help organize their design ideas or iterations into: home page, activity page, and post-activity page (Figure 3b). Similarly, participants were invited only to use the templates as optional support, and to create and ideate freely.

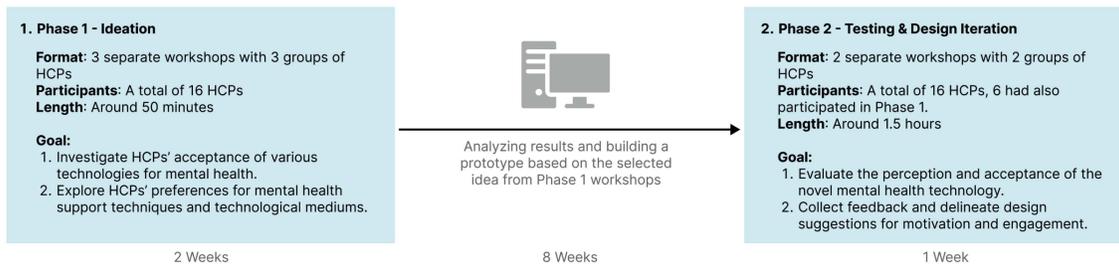

Figure 1: Overall research design and procedures

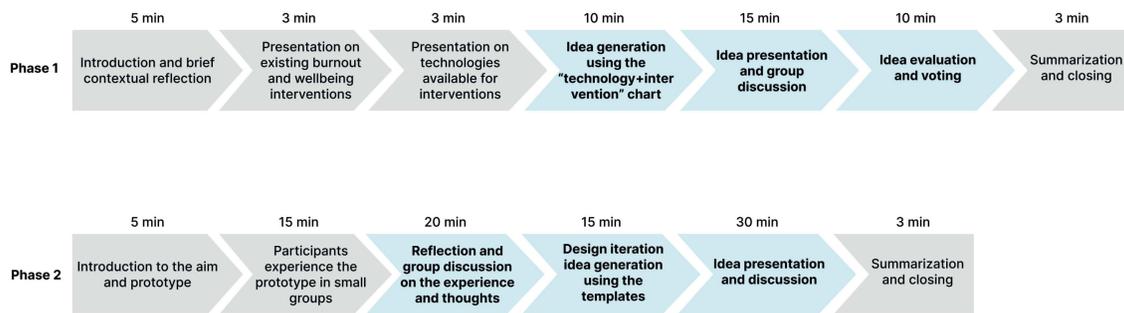

Figure 2: Overview of workshop flow

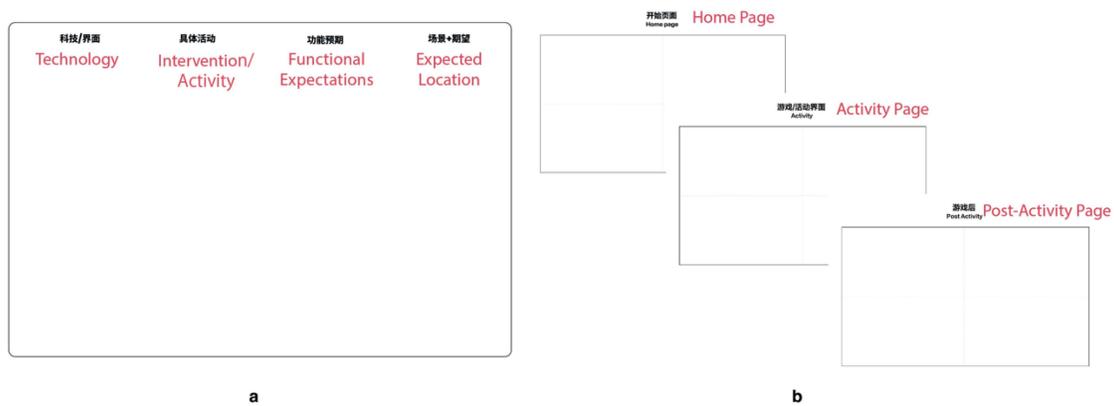

Figure 3: Design templates provided for participants for idea generation and design iteration

Table 1: Wellbeing technologies introduced to participants

| Technologies | Text introduction | Example applications in wellbeing interventions |
| --- | --- | --- |
| Virtual reality | A technology that immerses users in virtual environments through a head-mounted headset, allowing them to experience realistic 3D worlds. | 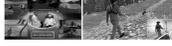 Ferrer Costa et al. [45] and Haliburton et al. [51] |
| Augmented reality | Technology that works to digitally overlay information and interactions, onto the real world through devices like smartphones or AR glasses. | 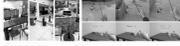 Woo et al. [123] and Ye et al. [126] |
| Embodied interaction | The use of physical movement and sensory input in technology and user interaction often engages user's body and mind and provides engaging experiences. | 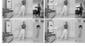 Huang et al. [58] |
| Tangible interface | Various technologies that allow users to interact with digital system through physical objects in order to provide a more hands-on experience. | 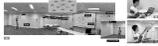 Bei et al. [13] |
| Ambient technology | Systems embedded in the environment that operate in the background, sensing and responding to users with light, sound and projections etc. | 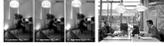 Yu et al. [128] and Chuang and Nieto [26] |
| Conversational agents | AI-driven systems, like chatbots, smart speakers or virtual assistants, interact with users through natural language, either via text or voice. | 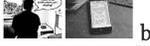 b Maharjan et al. [73] and Park et al. [85] |

## 3.2 Recruitment and Participants

The recruitment took place in Gansu Central Provincial Hospital, a tertiary hospital located in northwest China. The hospital is one of the largest in the region, with over 3,500 employees. The recruitment information was disseminated via email and direct messaging on WeChat [113]. Those who were interested in participation left their contact details and preferred time slots via an online questionnaire using an online survey platform in China called WJX [96]. Workshops were held in staff common rooms and a staff meeting room located within the hospital's main building. No monetary incentive was provided in this study, but food and beverages were provided to participants before and after each workshop session.

In total, 51 participants expressed interest in participating, and 25 dropped out due to timetable clashes or not responding to follow-up messages. Eventually, 26 participants were involved in the study (16 in Phase 1, 10 in Phase 2, and 6 in both Phases). Three workshops were held in Phase 1, each comprising 5, 7, and 4 participants. Although the research team tried to invite participants of Phase 1 to partake in the second phase, only 6 participants managed to attend due to their changing timetables. Therefore, the team invited new participants, and eventually another 10 participants attended Phase 2 workshops, which comprised 2 workshops, with 9 participants in the first and 7 in the second workshop. Participant details are listed in Table 2.

## 3.3 Data Collection, Analysis, and Ethics

All workshops in both phases were voice-recorded and transcribed for data analysis. Data in Phase 1 included co-design workshop transcripts, ideas generated by participants in the form of handwriting and sketches, and their preference voting. Data in Phase 2 included transcripts and design iteration in handwriting and sketches. Participants' handwriting and sketches were photographed and exported manually into PDF documents. The interviews and workshops were conducted in Mandarin Chinese.

Qualitative data were extracted from the transcription and analyzed using inductive thematic analysis [17]. Two authors (ZZ and JS) worked on the initial coding and theme development. The results were translated into English by two bilingual researchers (ZZ and

Table 2: Participant ID and demographics

| Participant ID | Age | Gender | Department | Job Type | Prior Technology Experiences* | Phases attended |
| --- | --- | --- | --- | --- | --- | --- |
| P1 | 28 | Female | Geriatrics | Physician | VR, AR, AT | 1 and 2 |
| P2 | 43 | Female | Geriatrics | Head nurse | AT, Wellbeing Apps | 1 and 2 |
| P3 | 39 | Female | Geriatrics | Nurse | CA, Wellbeing Apps | 1 |
| P4 | 26 | Female | Geriatrics | Nurse | AR, CA, AT, TEI | 1 |
| P5 | 47 | Male | Respiratory | Physician | VR, TEI | 1 |
| P6 | 32 | Female | Respiratory | Nurse | CA, AT | 1 |
| P7 | 35 | Female | General surgery | Nurse | Wellbeing Apps | 1 and 2 |
| P8 | 51 | Male | General surgery | Chief physician | CA, TEI | 1 and 2 |
| P9 | 32 | Male | General surgery | Physician | AR, AT, TEI | 1 |
| P10 | 40 | Male | Urology | Associate chief physician | CA | 1 |
| P11 | 39 | Male | Radiology | Radiologist | CA, Wellbeing Apps | 1 and 2 |
| P12 | 27 | Female | Radiology | Nurse | VR, CA | 1 and 2 |
| P13 | 24 | Female | Oncology | Medical trainee | CA, TEI | 1 |
| P14 | 24 | Female | Oncology | Medical trainee | AR, AT, Wellbeing Apps | 1 |
| P15 | 28 | Male | Oncology | Physician | TEI, CA | 1 |
| P16 | 44 | Male | Oncology | Associate chief physician | None | 1 |
| P17 | 35 | Female | Medical oncology | Physician | TEI, Wellbeing Apps | 2 |
| P18 | 36 | Female | Pediatrics | Head nurse | CA | 2 |
| P19 | 31 | Female | Chemotherapy | Nurse | AR, AT | 2 |
| P20 | 37 | Female | Chemotherapy | Head nurse | CA, Wellbeing Apps | 2 |
| P21 | 42 | Female | Cardiology | Associate chief physician | VR, CA | 2 |
| P22 | 37 | Female | Internal medicine | Head nurse | AT | 2 |
| P23 | 28 | Female | Internal medicine | Nurse | AR, CA | 2 |
| P24 | 28 | Male | Oncology | Physician | CA, TEI | 2 |
| P25 | 26 | Female | Oncology | Physician | VR, AT, Wellbeing Apps | 2 |
| P26 | 46 | Female | Oncology | Chief physician | CA | 2 |

*Note: AT = Ambient Technology, TEI = Tangible and Embodied Interface, CA = Conversational Agent

JS) and discussed among all authors. Initial findings, such as preliminary codes and themes, along with implications that emerged during the discussion, were used as a guide for the second round of coding. The second round was performed by authors (ZZ and JS), aiming to identify overlooked themes and refine the existing codes and themes. The outcome was again discussed and reviewed by all authors. NVivo 14 [69] (for Mac) was used to analyze the data. For both phases, we followed the same data analysis procedure.

The study was approved by the Gansu Provincial Hospital Research Ethics Committee and the Research Governance and Integrity Team of Imperial College London (Reference number: 22IC7585). A consent form was signed by each participant before the workshops. Participants were asked to read the participant information sheet, and were informed of the study purpose and content, and their rights to withdraw from the study at any time should they want to. The study was fully anonymized; no identifiable personal data were collected. Participants were explicitly informed of the anonymity of their responses and the measures taken to ensure their privacy and confidentiality.

## 4 Results

### 4.1 Phase 1: Staff visions of wellbeing technology in the hospital

In the first phase, participants were asked to create design ideas and then comment on and vote on ideas in each workshop session (each participant had three votes to distribute). After the 3 workshops, a total of 30 ideas were generated by participants (see Table 5 (Appendix B)). All ideas generated by participants were synthesized by the researchers into 7 final concepts (see Table 3), by combining similar concepts (except for 2 ideas, which were idiosyncratic and did not align with any of the rest). The preference votes for individual ideas were aggregated into the final synthesized concepts.

In the following paragraphs, we describe participants' key preferences and values embedded in the idea-generation workshops. All quotes from participants were followed by their participant ID.

*4.1.1 Potential novelty effect: HCPs' enthusiasm for alternative wellbeing technologies.* One of the notable findings from Phase 1 was the enthusiasm for wellbeing support delivered through embodied and immersive technology. Participants were mostly open and willing to try out alternative technologies regardless of whether they had encountered such technology in their lives previously.

Table 3: Synthesized design concepts, corresponding ideas and technologies involved in the concept

| Synthesized Concepts (aggregated preference votes) | Ideas Incorporated (see Appendix B) | Technology |
|---|---|---|
| **VR Venture** (13): A virtual reality intervention that allows HCPs to be transferred to different, non-clinical environments and experience activities like extreme sports, hiking, boxing etc. | Ideas 7, 15, 19, 20 and 26 | Virtual reality (VR) |
| **Sensory Nature** (6): A nature scape for mindfulness, relaxation and meditation in VR that recreates scenes (e.g. Chinese traditional gardens) and stimulates senses like audio, visual and smell. | Ideas 2, 11, and 20 | Virtual reality (VR) |
| **Digital Exercise Class** (10): Digital exercise, and mind-body movement (Yoga, Tai-chi or Qigong) class with colleagues, led by a virtual coach in a dedicated space within the hospital. | Ideas 1, 4, 5, 16, 22, and 29 | Embodied interaction |
| **Magic Mats** (8): Physical mats, or other tangible interfaces that can facilitate sports and exercise, like running, dancing, walking and boxing. | Ideas 9, 21, 26, and 30 | Tangible interface |
| **The Chameleon Room** (6): A staff room equipped with ambient technologies like projection mapping, soundscape and lighting that shifts to different environments for different wellbeing activities. | Ideas 3, 6, 12, 18, and 28 | Ambient technology |
| **Mind Graffiti** (3): An AR-based canvas, or physical wall with projection, for art therapy, where staff can have creative and expressive painting and doodling sessions individually or together with colleagues. | Ideas 10, 14, and 24 | Augmented reality (AR) |
| **TheraPal** (2): Conversational agents, in the form of chatbots or smart speakers, acting as a friend or tutor, for HCPs to provide psychotherapy, resilient training, and reliable advises. | Ideas 23 and 25 | Conversational agents |

For our participants, the attraction of varied technological experiences was a crucial motivator for engaging in wellbeing interventions. VR, embodied interaction using motion sensing, tangible interaction, and conversational agents were frequently mentioned by HCPs during idea generation. For example, P4, P5, and P6 were strong advocates for concepts around VR, and P5's quote below highlighted their motivation to engage with the technology:

> "I've experienced VR while traveling to different cities and have always been captivated by it - not just for its game elements, but for its ability to instantly transport you to another reality. I personally enjoy trying experiences I wouldn't dare attempt in real life, like skydiving. I believe VR would serve as an excellent relaxation tool for staff members here." (P5)

Participants' eagerness for new wellbeing technologies was consistent among both tech-savvy participants and those with less technology experience. For instance, P16, who just learned about motion sensing in the workshop, became quite excited about it and mentioned: "Oh, then if I want to practice Tai-chi, I can use motion sensing as well, right? ... I've always wanted to join a Tai-chi class but haven't had the chance." (P16).

*4.1.2 Technologies could improve clinical spaces and offer "transportive" experiences.* Another key aspect that independently arose across all 3 workshops in phase 1, is the potential of using technologies to transform the clinical environment or to "transport" HCPs into different environments, as reflected in P5's statement above about VRs' "ability to instantly transport you to another reality".

Firstly, participants mentioned ideas of using technologies like LED screens, projection mapping, and AR to transform the physical environment of their staff rooms and common rooms. According to them, most staff rooms in the hospital were underutilized or merely used as storage spaces. Participants P4, P9 and P11 described their ideas as being similar to "The Chameleon Room", envisioning how technologies could transform the current environment to create a valuable space for staff wellbeing:

> "Our staff room is small, and we have almost never used it for relaxation. We usually go to the canteen, or just stay at our own desks ... I wouldn't go to the staff room as there is nothing special about it ... if I can create a special staff room, with those immersive projections I saw in the shopping mall, and it shifts based on various occasions. Like, it can change to a yoga classroom or seaside scenes for meditation..." (P4)

Meanwhile, participants also praised the potential of technologies in terms of creating a "transportive" experience, one that allows them to detach from the clinical environment and mentally travel to another scenario, setting up a mental distance from the clinical environment even for a short time. This is reflected in the concepts of "VR Venture" and "Sensory Nature", in which staff expressed a need to be in another environment to "zone out" (P1) and "see other people's life and experience different stories" (P11). One participant well represented these feelings:

> "I vote for all technologies that can take me out of the hospital. No matter if it's about petting cats and dogs in the pet café or going to travel in nature ... you've got to get out of the working environment to relax, even if it's not real, like in a VR." (P6)

*4.1.3 Informal peer support and shared experiences.* In the workshops, participants highlighted the significance of social support and expressed interest in how technologies could facilitate peer connection and shared experiences among other participants.

As represented by the concepts of "Digital Exercise Class", "Magic Mats," and "Mind Graffiti", participants often mentioned a desire to have a collective experience with their colleagues. For example, P3 and P4 loved the idea of practicing yoga together:

> "We quite like yoga with music. My idea is, if there is a room with a yoga coach on TV, and motion sensing so I know how I did ... I wouldn't mind practicing it with my colleagues, actually that sounds pretty fun." (P3).

Similarly, P10, when commenting on the "Mind Graffiti" concept, mentioned:

> "We used to have this 'venting wall' when I was a trainee in [hospital name]. Within a small cohort, we would sometimes write thoughts on it, and there would be replies accumulating gradually. Nowadays, I find such elements are rare, and younger colleagues are always on their phones ... it would be helpful to build such connections within teams rather than only focusing on individuals" (P10)

One common characteristic of participants' preferences for social support was the informal or indirect approach. Unlike group-based interventions that directly focus on mental health topics, staff in this study emphasized informal peer support derived from collective experiences. For example, P8 mentioned his reluctance to talk about mental health openly with colleagues:

> "If it's doing mindfulness and Tai-chi, or exercise, etc. I'd like to join my colleagues and just have fun... but I don't think many people, especially us, the older generation, will enjoy talking openly about mental status, especially among those with whom you are not close." (P8)

*4.1.4 Tangible and embodied interaction facilitating mind-body skills.* Participants in all workshops generated ideas around physical exercise and mind-body practices facilitated by technologies. Participants regarded mobilizing and exercising routinely as an essential way to cope with work-related stress. For instance, P6 mentioned exercise and sports that can be assisted by technology:

> "Can we include ball games such as ping-pong or badminton in VR experiences? I find it always helpful to get active and exercise. These competitive sports are for off-work or longer breaks that we sometimes can have during a shift. For shorter breaks, it's helpful to do something less active, like jogging, just for your heart to relax ... you know, when we were kids, they had these dancing mats in the arcade?" (P6)

P6 then proposed her idea of "Magic Mats," which was an interactive mat that could be used to facilitate activities like dancing and running, with on-screen, light, and sound feedback. This idea was favored by many staff in the workshop and echoed similar proposals in the other two sessions (by P15 and P12). Many ideas center around tangible and embodied interaction using VR, motion sensing, and physical tools.

Another aspect staff brought up was fostering mind-body skills using embodied interaction. Participants across 3 workshops praised the positive effects of mind-body training and expressed interest in using technology to engage with such practices. For example, P1, P3, P4, P7, and P13 were keen to practice yoga, Tai-chi, or Qigong for wellbeing. They all proposed design ideas that are like the "Digital Exercise Class" concept and envisioned how technology could replace a human coach:

> "I know that in [another hospital], they have Yoga classes every week, led by professional yoga coaches. I was a bit jealous about it ... But if we can have this digital yoga coach on TV that guides your movements, and uses motion sensing, it would be quite nice as well." (P4).

*4.1.5 Technology reliability, infection control, and usability concerns.* Apart from the overall optimistic views on new forms of wellbeing-supportive technology, there were several participants who held reservations about the practicality of promoting digital interventions in the hospital. Their general concerns included the reliability of technologies in the long term, infection control concerns around tangible tools, lack of space for technology implementation, and potential usability issues.

For example, when discussing ideas around VR, P5 expressed concerns about the reliability of the technology and whether it can meet infection control requirements:

> "It's important to see how long the VR headset can run without being broken or malfunctioning. From my experience, these new technologies tend to break down after a while ... It's important to keep it sanitized as well, as we are in a hospital, so everything that touches staff and patient needs to meet strict infection control standards." (P5)

Similarly, P9 mentioned the importance of long-term stability of the technology and emphasized the importance of user-friendly design to avoid usability issues, especially for HCPs who are older and not tech-savvy.

## 4.2 Prototype development

Based on a combination of participant preference votes, results of the thematic analysis, and following discussion among the research team and clinical staff from the hospital, a set of concepts was identified, which could be fleshed out visually to gain further feedback from participants. Among these, the "Digital Exercise Class" was selected for development into a medium-fidelity prototype. This decision was made based on 1) "Digital Exercise Class" is the second-favorited concept by participants in Phase 1; 2) The concept matches the needs and values as revealed by the thematic analysis, especially the need for peer support and shared experiences, which is hard to be satisfied by concepts like "VR Venture."

The prototype was built in Unity3D version 2021.11. The prototype used a Microsoft Kinect 2.0 sensor for body tracking and was run on a Windows computer connected to a 75-inch screen. After carefully considering the types of mind-body movement, the research team selected the Qigong exercise, a mind-body exercise that is more suitable to practice within a limited space and is easier to learn for novice persons [63]. Before the activity, a short introduction to the mind-body movement technique (Qigong) was provided (see Figure 5a). During the activity, a Qigong instruction video that was publicly available online[1] was inserted to guide users' movement. Their body and movements were detected and reflected on the screen, represented by the default skeletons. Considering participants' preference for a more immersive experience, we decided to incorporate visual and sound elements of nature into the prototype. A seaside beach scene and sound effects were also incorporated into the prototype as a background (see Figure 5b). The prototype's other design and visual elements were kept minimal to give room for participant input.

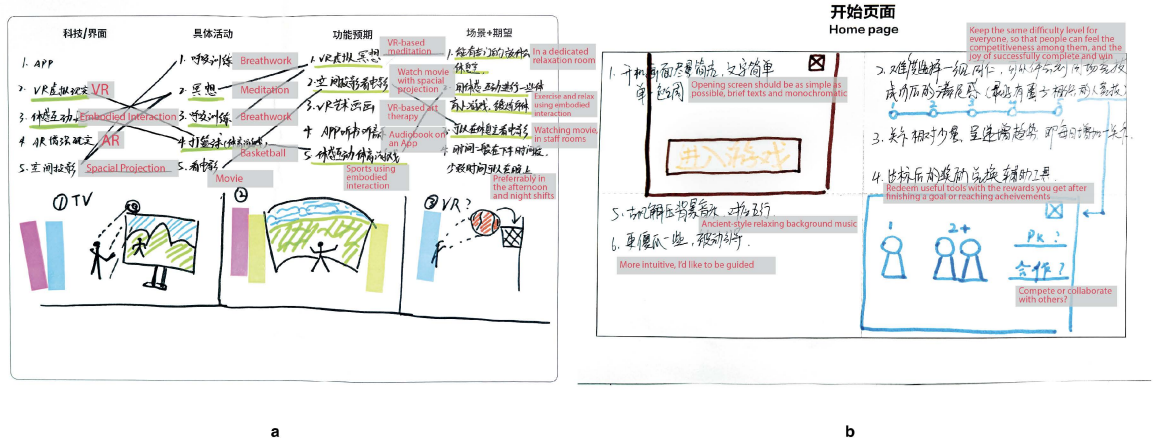

Figure 4: Participant handwriting and sketches in Phase 1 (a) and Phase 2 (b)

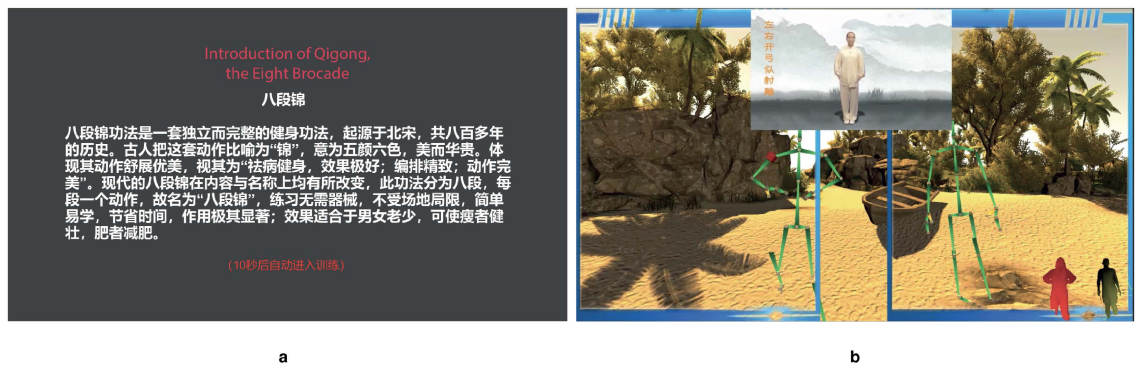

Figure 5: Screenshots of the medium-fidelity prototype

### 4.3 Prototype feedback and design improvement

All 16 participants tested the prototype and finished the 2-minute Qigong session. The Qigong sessions (Figure 6) were kept to approximately 2 minutes to accommodate all participants and reserve enough time for the design activities (as the participants had to take turns to experience the prototype). They were then invited to reflect on their experiences and ideate on how to improve the design of the prototype using the 3-stage design prompt provided. Their design feedback and reflections were summarized as themes below.

*4.3.1 Catering to the changing needs – Provide choices over activities, and engagement modes.* Participants demonstrated a clear need for choices in multiple aspects of an embodied wellbeing technology at work. Some commonly mentioned points include: 1) offering a variety of movement and exercise options with different levels of intensity, difficulty, and duration; 2) providing choices for single-player or multi-player modes; 3) enabling staff to choose between collaborative or competitive modes in multiplayer settings.

---
[1]https://www.bilibili.com/video/BV1gT4y1m7ec/

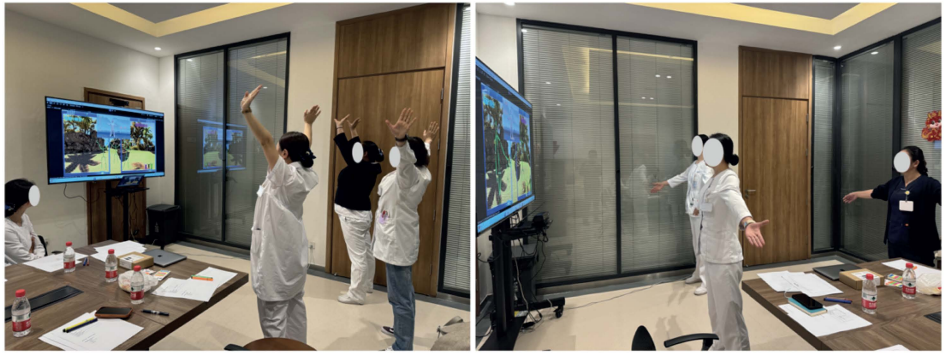

**Figure 6: Participants in the process of experiencing the prototype of Qigong exercise**

According to participants' feedback, HCPs' needs and preferences for wellbeing support could vary from time to time, based on their moods and sources of stress. For example, P22 suggested:

> "I'd split it into two modes. If I am feeling alright and would just like to chill and stretch my body, then Qigong and yoga would be nice. But sometimes, when I'm really stressed, or having a difficult patient, I'd choose something more intensive and sweating." (P22)

Similarly, P26 designed an assessment page for the system to recommend suitable activities:

> "I think there can be a page asking, are you feeling tired? Do you want to be on your own? Or what is your mental need at the moment? Then the system will recommend choices like either brief content or sweating aerobic activities." (P26).

Meanwhile, participants also highlighted the importance of catering to individual differences in personality and habits. For instance, P12 highlighted the personality differences and the impact on one's preference for specific activities and the choice between single-player or multi-player modes:

> "I think as HCPs, we are all different and we have our own ways of coping with stress. So, our needs for the digital tool are different ... it's helpful to have as many choices as possible for different staff. For example, for someone who is extrovert and active, it can offer collective and intense activities, and for people who are quieter like me, they can enjoy personal and less intense activities." (P12)

*4.3.2 Evidence fuels engagement – A need to see evidence of efficacy and personal improvements.* Receiving explicit feedback emerged as an important need for HCPs in the workshops. For example, staff preferred tracking stress and burnout levels to see how the measurements change after using the technology. P18 called it an "occupational habit to see numbers and figures." Another participant mentioned why it's crucial to provide detailed measurements and assessments:

> "I might be unaware of my actual stress level, and how it has changed before and after the activity ... I once saw a device in [another hospital], it shows how relaxed your muscles are on the computer ... sometimes people may feel physically relaxed, turns out their muscles are quite stiff and tense. And it's the same for your mind, you may feel very relaxed mentally, but your brain can be under strain." (P22)

Seeing the efficacy, benefits and their personal growth and improvement with clear evidence, such as changes in stress levels, was regarded as a key motivation to maintain staff engagement by several participants (P19, P20, and P25). These participants all designed around a post-activity feedback and analytics page. Similarly, other participants (P2 and P11) proposed using long- and short-term stress level reports, leveling-up challenges (P1, P17, P20 and P23), and positive reinforcement (P21) to foster a sense of growth and improvement.

*4.3.3 Quest for immersion into a non-clinical environment.* Echoing participants' desire for a "transportive experience" identified in Phase 1, we again found a common preference among participants for using digital interventions to immerse themselves in different environments. Participants (P1, P2, P11, P19 and P22 – P24) explicitly mentioned their preference for nature scenes while engaging in wellbeing interventions. For example, P11 praised the beach scene in the prototype, and emphasized the significance of being immersed into a relaxing environment:

> "I often practice along the exercise videos ... and I think it's crucial that the backdrop is a natural scene—just like the beach scene in game we just played, or a seaside, or even a mountain. Watching such scenery already creates a calming effect in my mind, following along with the exercise makes me even more relaxed." (P11)

Another participant applauded the benefits of on-screen or VR nature scenes, while also proposing her idea of "a character shift" from being an HCP into an anime character immersed in nature:

"I wonder if we could create an effect where a healthcare worker's white coat gradually fades away and disappears, revealing comfortable, casual clothing underneath. Then, we suddenly find ourselves in a vast, open field, evoking a feeling like a Hayao Miyazaki animation—romantic, ethereal, and deeply relaxing, immersing ourselves in nature." (P23)

Overall, similar to the feedback in Phase 1, participants expressed a strong desire to evade the clinical environments with the help of technology. As P26 concluded: "Try not to design any scenes or settings that relate to a hospital."

*4.3.4 Usability matters – the need for intuitive and effortless learning.* For many participants, usability is the primary consideration when using wellbeing technologies. To enhance the ease of use, especially where users are required to pay extra attention and learn new skills or knowledge, participants mentioned the importance of making the learning process as intuitive and effortless as possible. For example, for the activity introduction, participants displayed reluctance to read a large body of text, as shown in the prototype. Instead, they proposed ideas including replacing text with voiceover or short videos and animations like those on Douyin (TikTok) [21] (P8, P24 and P25). Moreover, P25 proposed the idea of using local dialects and accents in voiceover, which was applauded by many other staff.

Moreover, when it came to learning new skills, such as the Qigong movements demonstrated in the prototype, clear and immediate feedback and guidance were welcomed. Participants like P8, P11, P21, and P24 all expressed their preference for receiving clearer and more concise instructions on movement and breathing through voice, larger texts or colorful visual cues.

*4.3.5 Gamification can be facilitative but also confusing.* Most participants also highlighted the potential of gamification elements to create the experience of guidance and achievement. There were many ideas surrounding this theme, such as progress milestones, check-in challenges, follow-up tasks, and achievements and rewards.

For example, P18 expressed her dissatisfaction with the prototype, mentioning "a lack of guidance of continuous participation". She then designed a post-activity interface where there was a check-in challenge and follow-up tasks:

"[In her design] There are two kinds of guidance: one is in the short term, which is like a following-up request, such as asking if they want to play another round right now or later, or if they would like to listen to another audio as a kind of complementary task. Furthermore, you could encourage them to check in regularly in the long term. This kind of check-in encourages them to engage daily, and with each check-in, they are gradually guided towards better wellbeing." (P18)

Meanwhile, some participants debated the use of gamification elements like scores, rankings and rewards. P24, for example, mentioned his doubts about the scores while practicing Qigong, after another participant described his design of a scoring system: "With the scoring system, those who can do well in Tai-chi and Qigong will always score higher and remain in the top ranking". P22 also expressed objections to scoring in the activity on multiple fronts:

"It can be confusing that, if I scored low, does it only mean that I didn't do well in the movements? Or does it also imply that I get fewer health benefits from it?... I don't think a score is that relevant; even my son in primary school doesn't get exact scores in their exams anymore." (P22)

Participants (P1, P7, P18 – P20, P23 and P25) also discussed the idea of using virtual rewards to retain user engagement. Many of them highlighted the potential of using rewards such as positive reinforcement, complimentary items, and gifts that support the wellbeing of users, allowing them to better engage with the intervention. For instance, P1 and P19 mentioned "unlocking more avatars and their outfits", while P20 mentioned, "Earning coins and using them to purchase vouchers for private yoga classes or professional sessions like counseling and art therapy".

*4.3.6 It's more about the collective experiences.* Consistent with the findings of the concept generation workshops, participants generally appreciated the collective experience facilitated by the prototype. For a few like participants (P2, P18, and P26), although they had reservations about the intervention content, it was the collective experience that turned them into advocates of the prototype and the wellbeing technologies it signified:

"Our stress might come from work, interactions with patients, or even from personal life, especially for young colleagues ... and simply doing a few exercises might not be sufficiently effective ... I believe that, instead of focusing on stress reduction, you might as well just promote collective healthy activities. When people are together during the activity, it already lifts the mood. And this is especially true among HCPs ..." (P26)

Many other participants held similar views on the collective experience. P12 mentioned: "To me, it is motivating if I know there are many of us doing this together, and we can join the activity collectively." P21 regarded introducing and involving close colleagues in the activity as "quite a meaningful and fulfilling experience."

## 5 Discussion

To our knowledge, this is one of the first studies to probe the potential technologies that could support HCPs' wellbeing in a hospital setting. Through a co-design approach, our findings provided initial evidence on HCPs' perceptions and preferences for various technology-facilitated approaches to wellbeing support. Notably, although most digitally delivered workplace wellbeing interventions rely on personal devices such as smartphones and tablets [22, 48], our participants expressed a strong interest in alternative technologies.

The following recommendations present insights and design suggestions based on the findings of this study. Each one is elaborated on to include corroborating evidence from existing research, theoretical grounding, and implications for design.

## 5.1 Foster effective relaxation by providing immersive and "transportive" experiences

Previously, research on digital mental health has mainly focused on using personal technologies like smartphones, tablets and computers to provide wellbeing support for employee mental health, including for HCPs [1, 15]. Our findings demonstrate the potential and the advantages of using alternative technologies, such as VR, AR, and ambient technology, that support brief and effective breaks by transporting HCPs into relaxing environments.

Participants in both phases were enthusiastic about the idea of "transportive" experiences provided by immersive technologies. In the idea generation workshops, staff strongly preferred technologies that could mentally "transport" them from the clinical environment (concept: VR Venture, Sensory Nature, and The Chameleon Room). Such preferences for a transportive experience were also commonly shared in Phase 2, as participants praised the potential to immerse themselves in nature through head-mounted devices or large screens. Indeed, for HCPs worldwide, the clinical environment is recognized as a major source of work-related stress [119]. Digital relaxation interventions using head-mounted devices, projections, and large display systems offer a practical antidote by facilitating a mental break from the environment [2, 45]. For example, recent studies have shown preliminary results of short experiences in nature using VR- or high-definition screens, which promoted positive emotions and reduced stress and anxiety among HCPs [2, 14, 127]. Other studies also suggested that immersive and transportive experiences contributed to user retention in digital mental health interventions [45, 86]. Hence, we suggest future work to further explore and evaluate technology-facilitated transportive experiences as mental health support for HCPs.

However, it is important to recognize that an exclusive or excessive focus on such transportive experiences and mental escape may inadvertently lead to detrimental outcomes, including avoidance coping mechanisms. Avoidance coping, as one of the three common coping mechanisms among HCPs facing stress and burnout [74, 89], involves avoiding the stressors by focusing on distractions rather than managing them directly [74]. It is, therefore, less effective in preventing burnout and may even exacerbate it, leading to deepened depersonalization and emotional exhaustion [98]. Hence, we suggest that immersive and transportive technologies be utilized strategically—either integrated with structured mental health support and psychoeducational content, or offered for short restorative breaks. Additionally, immersive technologies can serve as a gateway or catalyst to enhance HCPs' engagement with traditional in-person interventions requiring sustained engagement. We also suggest HCI and mental health researchers integrate immersive and transportive elements into other digital interventions to support both immediate relaxation and long-term engagement and wellbeing outcomes.

## 5.2 The value of embodied and on-site wellbeing interventions

Embodied interaction was highly appraised by HCPs in both phases of the study. During the idea generation workshops, a large part of the conversation revolved around ideas for using technologies to engage with physical movements and mind-body practices in the workplace (i.e., the Digital Exercise Class and Magic Mats concepts). Furthermore, during the testing workshop, the Qigong exercise prototype was well-received.

Indeed, as HCPs often suffer from physical pain along with psychological strain [75], promoting mind-body movement and brief exercises during breaks or between shifts can be especially valuable in mitigating stress, burnout, and anxiety at work [27, 46, 63, 74]. Regarding embodied exercises assisted by technologies, recent studies in HCI and mental health have shown preliminary evidence of their efficacy on staff wellbeing and stress reduction [46, 49]. We suggest future studies further explore the potential of embodied technology in diverse healthcare settings, by probing into the long-term effects, and providing guidelines on the design and delivery of such interventions to maximize user engagement and wellbeing benefits.

Our findings also highlight the importance of offering on-site wellbeing interventions for HCPs, who are arguably less affected by the remote working trend. Participants in both phases were especially keen on interventions using VR, ambient intelligence, and embodied technologies directly available in their workplace. While on-site technologies may not be as convenient to use as mental health apps or websites, they have unique affordances that may enhance user engagement. Firstly, on-site interventions can be seamlessly integrated into the working environment and HCPs' daily routines. In the current study, participants envisioned that an intervention could be "applied in the staff room" (P3) and "used right after the morning debrief to energize you and get you ready for the day" (P25). Moreover, the physical embodiment of these interventions can serve as a contextual reminder. This presents an opportunity to naturally overcome forgetfulness, which is a crucial barrier to user engagement with mental health apps and websites [16, 40, 60]. A visual cue also avoids reliance on disruptive notifications, which research has shown is often ineffective in sustaining engagement [57]. Placing on-site and embodied wellbeing technologies in high-traffic areas of hospitals, such as staff rooms, common areas, and canteens, provides a constant, tangible prompt for engagement. Therefore, we propose the application of on-site interventions for populations like HCPs, who have extremely busy and unpredictable schedules but who spend many hours on-site. We suggest future studies to test and provide initial evidence on the potential of embodied and on-site wellbeing technologies to improve user engagement among HCPs.

## 5.3 Engagement beyond the novelty effect – satisfying the needs of autonomy, competence, and relatedness

Our results indicated that HCPs' were generally open and keen on novel and alternative technologies for wellbeing support. However, such enthusiasm may partly reflect the novelty effect, which facilitates user engagement in the short term and fades with repeated exposures [109].

To address the challenge of sustaining motivation and engagement beyond the novelty effect, Self-Determination Theory (SDT) offers a valuable lens [38]. SDT is one of the most widely used and validated theories on human motivation and wellbeing [7, 93], and has been utilized by many HCI researchers to study user motivation and engagement with technologies and products, such as games [103, 118], social media [19], and digital health platforms [93, 94]. SDT proposes that three basic psychological needs have to be met within the user experience to support user wellbeing, motivation, and long-term engagement, which are autonomy (a feeling of volition, self-endorsement, and acting in accordance with one's values and goals), competence (feeling effective and capable) and relatedness (a feeling of meaningful connection with others) [101, 102]. These three basic psychological needs have been validated across countries and cultures and are regarded as universal to human motivation and across domains, including within user interaction with technologies [23, 37, 91].

In this study, HCPs' feedback and reflections on the prototype have provided several insights, echoing the basic psychological need theory in SDT. Specifically, our findings indicated that HCPs highly valued wellbeing technologies that supported their user autonomy, competence, and relatedness in the user experience. We provide design implications based on our findings, through an SDT lens and supported by SDT-related evidence, in the following sections.

*5.3.1 Support Autonomy by offering room for choices and supporting control over mental health.* As one of the fundamental psychological human needs theorized by SDT, autonomy is recognized as the key to user motivation and engagement with technologies [91]. In this study, the need for autonomy was also a recurring theme in participants' feedback and design ideations. For example, providing sufficient choices to meet their individually different and ever-shifting wellbeing needs was key for many participants. HCPs highlighted the need for choices in almost every part of user experience, from personalizing intervention content, and selecting personal, collaborative, or competitive modes, to customizing avatars and visual styles. Providing users with adequate choices was also emphasized in previous studies to support user autonomy [18, 55, 88]. For example, Harjumaa et al. described an application designed to support behavioral change to prevent type-2 diabetes, which offered a broad selection of behaviors to foster user autonomy [55].

Supporting HCPs' control over their mental health can also foster their sense of autonomy, as highlighted by our findings. First, participants strongly preferred wellbeing support that was *informal (i.e., not officially a "mental health" intervention),* which they described as "stress-free." They also expressed reservations about activities that impose pressure to disclose mental health concerns. This can be a result of higher stigma around mental health in healthcare environments and in societies where stigma is more pervasive [64, 87]. Hence, it is important to provide a stress-free experience, where mental health interventions are not formally labeled as such and where disclosure is controlled by themselves and not enforced by the intervention.

On the other hand, participants also emphasized the importance of receiving feedback from technology, specifically in the form of concrete numbers and figures. This inclination toward clear, quantifiable feedback represented their desire for a sense of control and in-depth understanding in relation to their wellbeing. Previous studies have also suggested providing quantitative feedback to support user autonomy and engagement [97]. For HCPs, we argue that it may be even more valuable to offer measurable outcomes, as they work in an environment where evidence and health data guide decision-making. Providing feedback and wellbeing data comprehensively aligns with their professional mindset, provides a meaningful rationale for sustained engagement, and empowers them to make informed choices, all of which are fundamental to their autonomy [90].

*5.3.2 Support competence by ensuring usability and clarity, and fostering personal improvement.* Usability is a primary requirement for ensuring users' sense of competence in a technology experience [92, 112]. It is known that usability barriers are a major cause of user attrition [16, 47]. Within hospital settings, novel technologies such as VR, embodied systems, and ambient intelligence introduce unique usability challenges. As shown in our findings, participants were worried about the robustness and durability of the new wellbeing technologies and the potential costs to maintain them under strict infection control standards. Indeed, implementing and managing wellbeing interventions in the hospital can be a complex task that requires collaboration among many stakeholders, including hospital administration, clinical staff, facilities management, infection control personnel, and the environment service team, to ensure practical efficacy and usability. Thinking comprehensively about ongoing maintenance can effectively facilitate a successful implementation of wellbeing technologies in complex environments like hospitals.

On the other hand, common design and interface elements that impact usability [34] were also highlighted by HCPs in this study, including their preference for clearer and shorter introductions, clarity on gamification elements, and the use of audio-visual information to enable an intuitive learning experience. For HCPs, as also implied in previous studies, it is important for new technological interventions to minimize the additional cognitive burden to prevent adding to their existing mental load [61, 79]. Moreover, gamification elements, which are now commonly regarded as an engagement-facilitating approach in HCI [6, 13, 24], may inadvertently diminish users' competence if not applied with caution [90]. Participants in our study demonstrated a preference for positive and non-judgmental feedback, and a clearer rationale for gamification elements like scores and rewards. Indeed, this aligns with SDT-based design heuristics, which recommend non-evaluative and effectance-relevant feedback to support user competence [90].

Participants also highlighted the importance of experiencing a sense of progress and improvement to sustain their motivation. Echoing previous research evidence, providing informational feedback with evidence of progress and personal improvement can effectively support users' need for competence and sustain engagement beyond the novelty effect [90, 109]. HCPs in this study suggested regular check-ins and follow-up tasks, providing clear guidance over time, and demonstrating personal improvement using clear longitudinal data and reports, which can support wellbeing interventions to cultivate a sense of growth and personal competence. In sum, we suggest future studies in HCI and mental health be mindful of gamification elements and styles of feedback and ensure these align with research-based guidelines for supporting autonomy and competence [25, 90].

*5.3.3 Promote connectedness by facilitating meaningful collective experiences.* According to previous literature, many traditional wellbeing interventions for HCPs incorporate social support through group-based activities [84]. This allows staff to benefit not only from the interventions but also from the social connections fostered through shared experiences [4, 122]. Compared to other employees, healthcare professionals especially rely on connectedness among peers for their resilience and wellbeing [67]. For example, studies showed that camaraderie and social support were crucial for HCPs' mental health during and after COVID-19 [70, 106, 108, 124]. However, such social elements and the importance of forming therapeutic alliances between colleagues were often overlooked in the existing literature on digital mental health interventions [67].

Our findings identified various ways novel technologies could facilitate connectedness and meaningful shared experiences for HCPs. For example, in Phase 1, many participant-generated ideas involved shared experiences (i.e., Digital Exercise Class and Mind Graffiti concepts). These ideas highlighted the importance of informal social support derived from collective experiences, echoing findings and implications of a recent review [119]. Furthermore, in the second phase, the collective learning experience facilitated by the prototype was also valued by participants—many even emphasized that the benefits of a collective experience outweighed the significance of the intervention itself. We suggest that by promoting meaningful connection and support among HCPs, wellbeing interventions can be more engaging and generate better outcomes. Similar recommendations were also proposed by previous studies in HCI, where social support was emphasized for engaging and effective digital mental health interventions [60, 110]. Combining our findings, we recommend that future studies probe into the potential of facilitating peer connection through shared experiences, by using technologies that allow collective engagement and interaction. This enables digital wellbeing interventions to bring back the social element that played an indispensable role in traditional interventions. Supporting connectedness can also fuel long-term engagement, as previous studies identified the importance of social motivations in overcoming the novelty effect and sustaining user engagement [76, 109].

It's worth noting that, although the collectivist culture in China could affect participants' inclination toward peer support and connection, multiple studies have provided evidence of the importance of social support and connectedness among HCPs across individualist and collectivist cultures [41, 119]. Moreover, as shown by SDT, relatedness is a universal and fundamental psychological need [37, 100]. Hence, we argue that providing social features and facilitating peer-to-peer alliances using technologies are worth exploring in future studies. Nonetheless, the collective interaction should be made optional, as our findings also demonstrated that HCPs' wellbeing needs vary over time and from person to person. It is important to provide the choice of shared experiences while allowing HCPs to freely choose from different modes of engagement. This could safeguard the wellbeing technologies to satisfy users' need for relatedness in an autonomy-supportive way, echoing design implications and heuristics in previous studies [19, 90].

In summary, the above recommendations, drawn from our findings, are consistent with both existing research and with the principles of, and heuristics derived from, self-determination theory.

Framing the findings in this light can help with understanding the root causes of these experiences as well as how to effectively address them in future technology designs for workplace wellbeing in hospitals.

## 5.4 Limitations and Future Research

One limitation of this study is the participant sample, which is comprised of HCPs (mostly female) in a single hospital in China. Workplace and cultural differences can impact participants' views and opinions. However, we argue that there are still considerable similarities among HCPs globally, due to the similar nature of their roles and the challenges they encounter regularly. Regardless of their location and cultural backgrounds, most HCPs worldwide share common ground in their commitment to patient care, evidence-based practices, and the stressors and contingencies they face [70, 99]. Recent research has shown similarity in the healthcare workforce and their wellbeing needs across the West and the East [129]. Hence, our findings and implications in this study could be extended to other cultural contexts. Nonetheless, we recommend future studies to further validate and refine the preferences and values identified in this study, by conducting similar workshops with HCPs of different socio-cultural backgrounds.

Additionally, the use of a volunteer sample composed solely of end users (HCPs) represents another limitation of this study. Specifically, the lack of involvement of other stakeholders, such as hospital administrators and managers, could cause the underrepresentation of different perspectives. Plus, it introduces the potential for participant self-selection bias, where HCPs who were willing to participate may already have experience with or interest in alternative or digital technologies. Hence, we suggest future co-design studies involve more diverse, randomized samples of stakeholders in the healthcare setting to mitigate potential biases and ensure the co-design findings incorporate diverse perspectives.

This study is a preliminary work that focuses on co-design with HCPs and investigates their perspectives and preferences. Hence, no quantitative measures were introduced in the study to test the acceptability or stress and burnout outcomes. Future studies should focus on pilot testing the prototypes in real healthcare settings to determine acceptability, feasibility, and potential effectiveness on HCPs' mental wellbeing.

In addition, although our data showed participants were highly enthusiastic about alternative technologies, appreciating their immersive and embodied affordances, some of this enthusiasm could arise from a novelty effect, as these technologies are still relatively new in the mainstream digital mental health field. However, we suggest that the incorporation of motivational theories like SDT could help derive design recommendations that overcome novelty effects and facilitate engagement in the long term. Longitudinal work would be necessary to evaluate the technologies and design recommendations we provide and explore how user engagement changes over time.

Another limitation of this study stems from the limited access to prototypes in testing workshops. Due to time and technical constraints, more technology-based interventions were not developed and tested among HCPs. We recommend future studies to test and evaluate other forms of technology, with content tailored for HCPs that supports their needs for transportive, embodied and collective experiences for wellbeing.

## 6 Conclusion

Based on a 2-phase co-design approach, this study investigated HCPs' needs and preferences for wellbeing interventions with a focus on alternative forms of technology. We identified promising directions for technologies to support the wellbeing of HCPs who preferred transportive, embodied and collective experiences. We also draw connections between findings and how interventions can be designed to support basic psychological needs (i.e., autonomy, competence, and relatedness). We argue that, by providing the experiences that are valued and needed by HCPs in a way that supports basic psychological need satisfaction, future studies in HCI and mental health are more likely to develop digital interventions that overcome the novelty effect and engage HCPs in the long term. Additionally, we advocate for more studies to explore how alternative technologies can connect with, or complement, mental health apps and in-person interventions. These technologies could serve as an engaging gateway for users to familiarize themselves with mental health support while also providing immersive and collective experiences that sustain user engagement and broaden accessibility to mental health support.

## Acknowledgments

This study is sponsored by CW+ and the Chinese Scholarship Council. We would like to thank all the HCPs who participated in the study.

## References

[1] Daniela Adam, Julia Berschick, Julia K. Schiele, Martin Bogdanski, Marleen Schröter, Melanie Steinmetz, Anna K. Koch, Jalid Sehouli, Sylvia Reschke, Wiebke Stritter, Christian S. Kessler, and Georg Seifert. 2023. Interventions to reduce stress and prevent burnout in healthcare professionals supported by digital applications: a scoping review. *Frontiers in Public Health* 11: 1231266. https://doi.org/10.3389/fpubh.2023.1231266

[2] Jai Shree Adhyaru and Charlotte Kemp. 2022. Virtual reality as a tool to promote wellbeing in the workplace. *DIGITAL HEALTH* 8: 205520762210844. https://doi.org/10.1177/20552076221084473

[3] Pasquale Arpaia, Giovanni D'Errico, Lucio Tommaso De Paolis, Nicola Moccaldi, and Fabiana Nuccetelli. 2022. A Narrative Review of Mindfulness-Based Interventions Using Virtual Reality. *Mindfulness* 13, 3: 556–571. https://doi.org/10.1007/s12671-021-01783-6

[4] Wendy L. Awa, Martina Plaumann, and Ulla Walter. 2010. Burnout prevention: A review of intervention programs. *Patient Education and Counseling* 78, 2: 184–190. https://doi.org/10.1016/j.pec.2009.04.008

[5] Luke Balcombe and Diego De Leo. 2021. Digital Mental Health Challenges and the Horizon Ahead for Solutions. *JMIR Mental Health* 8, 3: e26811. https://doi.org/10.2196/26811

[6] Nick Ballou, Sebastian Deterding, Ioanna Iacovides, and Laura Helsby. 2022. Do People Use Games to Compensate for Psychological Needs During Crises? A Mixed-Methods Study of Gaming During COVID-19 Lockdowns. In *CHI Conference on Human Factors in Computing Systems*, 1–15. https://doi.org/10.1145/3491102.3501858

[7] Nick Ballou, Sebastian Deterding, April Tyack, Elisa D Mekler, Rafael A Calvo, Dorian Peters, Gabriela Villalobos-Zúñiga, and Selen Turkay. 2022. Self-Determination Theory in HCI: Shaping a Research Agenda. In *CHI Conference on Human Factors in Computing Systems Extended Abstracts*, 1–6. https://doi.org/10.1145/3491101.3503702

[8] Rosa Mª Baños, Rocío Herrero, and Mª Dolores Vara. 2022. What is the Current and Future Status of Digital Mental Health Interventions? *The Spanish Journal of Psychology* 25: e5. https://doi.org/10.1017/SJP.2022.2

[9] Kate Barrett and Ian Stewart. 2021. A preliminary comparison of the efficacy of online Acceptance and Commitment Therapy (ACT) and Cognitive Behavioural Therapy (CBT) stress management interventions for social and healthcare workers. *Health & Social Care in the Community* 29, 1: 113–126. https://doi.org/10.1111/hsc.13074

[10] Belén Barros Pena, Nelya Koteyko, Martine Van Driel, Andrea Delgado, and John Vines. 2023. "My Perfect Platform Would Be Telepathy" - Reimagining the Design of Social Media with Autistic Adults. In *Proceedings of the 2023 CHI Conference on Human Factors in Computing Systems*, 1–16. https://doi.org/10.1145/3544548.3580673

[11] Gil Bar-Sela, Doron Lulav-Grinwald, and Inbal Mitnik. 2012. "Balint Group" Meetings for Oncology Residents as a Tool to Improve Therapeutic Communication Skills and Reduce Burnout Level. *Journal of Cancer Education* 27, 4: 786–789. https://doi.org/10.1007/s13187-012-0407-3

[12] Amit Baumel, Frederick Muench, Stav Edan, and John M Kane. 2019. Objective User Engagement With Mental Health Apps: Systematic Search and Panel-Based Usage Analysis. *Journal of Medical Internet Research* 21, 9: e14567. https://doi.org/10.2196/14567

[13] Rongqi Bei, Yixuan Li, and Xin Tong. 2021. Whack-a-Ball: An Exergame Exploring the Use of a Ball Interface for Facilitating Physical Activities. In *Extended Abstracts of the 2021 Annual Symposium on Computer-Human Interaction in Play*, 249–255. https://doi.org/10.1145/3450337.3483465

[14] Elizabeth Beverly, Laurie Hommema, Kara Coates, Gary Duncan, Brad Gable, Thomas Gutman, Matthew Love, Carrie Love, Michelle Pershing, and Nancy Stevens. 2022. A tranquil virtual reality experience to reduce subjective stress among COVID-19 frontline healthcare workers. *PLOS ONE* 17, 2: e0262703. https://doi.org/10.1371/journal.pone.0262703

[15] Holly Blake, Fiona Bermingham, Graham Johnson, and Andrew Tabner. 2020. Mitigating the Psychological Impact of COVID-19 on Healthcare Workers: A Digital Learning Package. *International Journal of Environmental Research and Public Health* 17, 9: 2997. https://doi.org/10.3390/ijerph17092997

[16] Judith Borghouts, Elizabeth Eikey, Gloria Mark, Cinthia De Leon, Stephen M Schueller, Margaret Schneider, Nicole Stadnick, Kai Zheng, Dana Mukamel, and Dara H Sorkin. 2021. Barriers to and Facilitators of User Engagement With Digital Mental Health Interventions: Systematic Review. *Journal of Medical Internet Research* 23, 3: e24387. https://doi.org/10.2196/24387

[17] Virginia Braun and Victoria Clarke. 2006. Using thematic analysis in psychology. *Qualitative Research in Psychology* 3, 2: 77–101. https://doi.org/10.1191/1478088706qp063oa

[18] Luke Brownlow. 2022. Targeting the Needs of Self-Determination Theory: An Overview of Mental Health Care Apps. *European Journal of Mental Health* 17, 1: 91–100. https://doi.org/10.5708/EJMH/17.2022.1.8

[19] Ryan Burnell, Dorian Peters, Richard M. Ryan, and Rafael A. Calvo. 2023. Technology evaluations are associated with psychological need satisfaction across different spheres of experience: an application of the METUX scales. *Frontiers in Psychology* 14: 1092288. https://doi.org/10.3389/fpsyg.2023.1092288

[20] Amy Burton, Catherine Burgess, Sarah Dean, Gina Z. Koutsopoulou, and Siobhan Hugh-Jones. 2017. How Effective are Mindfulness-Based Interventions for Reducing Stress Among Healthcare Professionals? A Systematic Review and Meta-Analysis: Mindfulness Interventions for Stress Reduction. *Stress and Health* 33, 1: 3–13. https://doi.org/10.1002/smi.2673

[21] Bytedance. TikTok. Retrieved from https://www.tiktok.com/en/

[22] Julianna Catania, Steph Beaver, Rakshitha S Kamath, Emma Worthington, Minyi Lu, Hema Gandhi, Heidi C Waters, and Daniel C Malone. 2024. Evaluation of Digital Mental Health Technologies in the United States: Systematic Literature Review and Framework Synthesis. *JMIR Mental Health* 11: e57401. https://doi.org/10.2196/57401

[23] Beiwen Chen, Maarten Vansteenkiste, Wim Beyers, Liesbet Boone, Edward L. Deci, Jolene Van Der Kaap-Deeder, Bart Duriez, Willy Lens, Lennia Matos, Athanasios Mouratidis, Richard M. Ryan, Kennon M. Sheldon, Bart Soenens, Stijn Van Petegem, and Joke Verstuyf. 2015. Basic psychological need satisfaction, need frustration, and need strength across four cultures. *Motivation and Emotion* 39, 2: 216–236. https://doi.org/10.1007/s11031-014-9450-1

[24] Vanessa Wan Sze Cheng, Tracey A Davenport, Daniel Johnson, Kellie Vella, Jo Mitchell, and Ian B Hickie. 2018. An App That Incorporates Gamification, Mini-Games, and Social Connection to Improve Men's Mental Health and Well-Being (MindMax): Participatory Design Process. *JMIR Mental Health* 5, 4: e11068. https://doi.org/10.2196/11068

[25] Vanessa Wan Sze Cheng, Sarah E Piper, Antonia Ottavio, Tracey A Davenport, and Ian B Hickie. 2021. Recommendations for Designing Health Information Technologies for Mental Health Drawn From Self-Determination Theory and Co-design With Culturally Diverse Populations: Template Analysis. *Journal of Medical Internet Research* 23, 2: e23502. https://doi.org/10.2196/23502

[26] Yaliang Chuang and Jose E Gallegos Nieto. 2022. Element: An Ambient Display System for Evoking Self-Reflections and Supporting Social-Interactions in a Workspace. In *CHI Conference on Human Factors in Computing Systems Extended Abstracts*, 1–7. https://doi.org/10.1145/3491101.3519721

[27] Rosario Cocchiara, Margherita Peruzzo, Alice Mannocci, Livia Ottolenghi, Paolo Villari, Antonella Polimeni, Fabrizio Guerra, and Giuseppe La Torre. 2019. The Use of Yoga to Manage Stress and Burnout in Healthcare Workers: A Systematic Review. *Journal of Clinical Medicine* 8, 3: 284. https://doi.org/10.3390/jcm8030284

[28] Karen Cochrane, Lian Loke, Andrew Campbell, and Naseem Ahmadpour. 2020. Mediscape: Preliminary Design Guidelines for Interactive Rhythmic Soundscapes for Entraining Novice Mindfulness Meditators. In *32nd Australian Conference on Human-Computer Interaction*, 379–391. https://doi.org/10.1145/3441000.3441052

[29] Catherine Cohen, Silvia Pignata, Eva Bezak, Mark Tie, and Jessie Childs. 2023. Workplace interventions to improve well-being and reduce burnout for nurses, physicians and allied healthcare professionals: a systematic review. *BMJ Open* 13, 6: e071203. https://doi.org/10.1136/bmjopen-2022-071203

[30] Ruth E. Cooper, Katherine R. K. Saunders, Anna Greenburgh, Prisha Shah, Rebecca Appleton, Karen Machin, Tamar Jeynes, Phoebe Barnett, Sophie M. Allan, Jessica Griffiths, Ruth Stuart, Lizzie Mitchell, Beverley Chipp, Stephen Jeffreys, Brynmor Lloyd-Evans, Alan Simpson, and Sonia Johnson. 2024. The effectiveness, implementation, and experiences of peer support approaches for mental health: a systematic umbrella review. *BMC Medicine* 22, 1: 72. https://doi.org/10.1186/s12916-024-03260-y

[31] Shane P. Cross, Eyal Karin, Lauren G. Staples, Madelyne A. Bisby, Katie Ryan, Georgia Duke, Olav Nielssen, Rony Kayrouz, Alana Fisher, Blake F. Dear, and Nickolai Titov. 2022. Factors associated with treatment uptake, completion, and subsequent symptom improvement in a national digital mental health service. *Internet Interventions* 27: 100506. https://doi.org/10.1016/j.invent.2022.100506

[32] Timothy Culbert. 2017. Perspectives on Technology-Assisted Relaxation Approaches to Support Mind-Body Skills Practice in Children and Teens: Clinical Experience and Commentary. *Children* 4, 4: 20. https://doi.org/10.3390/children4040020

[33] Claudia Daudén Roquet, Nikki Theofanopoulou, Jaimie L Freeman, Jessica Schleider, James J Gross, Katie Davis, Ellen Townsend, and Petr Slovak. 2022. Exploring Situated & Embodied Support for Youth's Mental Health: Design Opportunities for Interactive Tangible Device. In *CHI Conference on Human Factors in Computing Systems*, 1–16. https://doi.org/10.1145/3491102.3502135

[34] Marco De Angelis, Lucia Volpi, Davide Giusino, Luca Pietrantoni, and Federico Fraboni. 2024. Acceptability and Usability of a Digital Platform Promoting Mental Health at Work: A Qualitative Evaluation. *International Journal of Human–Computer Interaction*: 1–14. https://doi.org/10.1080/10447318.2024.2313892

[35] Stefan De Hert. 2020. Burnout in Healthcare Workers: Prevalence, Impact and Preventative Strategies. *Local and Regional Anesthesia* Volume 13: 171–183. https://doi.org/10.2147/LRA.S240564

[36] M Deady, D Peters, H Lang, R Calvo, N Glozier, H Christensen, and S B Harvey. 2017. Designing smartphone mental health applications for emergency service workers. *Occupational Medicine* 67, 6: 425–428. https://doi.org/10.1093/occmed/kqx056

[37] Edward L. Deci, Richard M. Ryan, Marylène Gagné, Dean R. Leone, Julian Usunov, and Boyanka P. Kornazheva. 2001. Need Satisfaction, Motivation, and Well-Being in the Work Organizations of a Former Eastern Bloc Country: A Cross-Cultural Study of Self-Determination. *Personality and Social Psychology Bulletin* 27, 8: 930–942. https://doi.org/10.1177/0146167201278002

[38] Kevin Doherty and Gavin Doherty. 2019. Engagement in HCI: Conception, Theory and Measurement. *ACM Computing Surveys* 51, 5: 1–39. https://doi.org/10.1145/3234149

[39] Nina Döllinger, David Mal, Sebastian Keppler, Erik Wolf, Mario Botsch, Johann Habakuk Israel, Marc Erich Latoschik, and Carolin Wienrich. 2024. Virtual Body Swapping: A VR-Based Approach to Embodied Third-Person Self-Processing in Mind-Body Therapy. In *Proceedings of the CHI Conference on Human Factors in Computing Systems*, 1–18. https://doi.org/10.1145/3613904.3642328

[40] Liesje Donkin and Nick Glozier. 2012. Motivators and Motivations to Persist With Online Psychological Interventions: A Qualitative Study of Treatment Completers. *Journal of Medical Internet Research* 14, 3: e91. https://doi.org/10.2196/jmir.2100

[41] Jia Fan, Yuyang Chang, Li Li, Nan Jiang, Zhifei Qu, Jiaxin Zhang, Meihua Li, Bing Liang, and Danhua Qu. 2024. The relationship between medical staff burnout and subjective wellbeing: the chain mediating role of psychological capital and perceived social support. *Frontiers in Public Health* 12: 1408006. https://doi.org/10.3389/fpubh.2024.1408006

[42] Ronghua Fang and Xia Li. 2015. A regular yoga intervention for staff nurse sleep quality and work stress: a randomised controlled trial. *Journal of Clinical Nursing* 24, 23–24: 3374–3379. https://doi.org/10.1111/jocn.12983

[43] Alexz Farrall, Jordan Taylor, Ben Ainsworth, and Jason Alexander. 2023. Manifesting Breath: Empirical Evidence for the Integration of Shape-changing Biofeedback-based Artefacts within Digital Mental Health Interventions. In *Proceedings of the 2023 CHI Conference on Human Factors in Computing Systems*, 1–14. https://doi.org/10.1145/3544548.3581188

[44] Jose Ferrer Costa, Nuria Moran, Carlos Garcia Marti, Leomar Javier Colmenares Hernandez, Florin Radu Ciorba Ciorba, and Maria Jose Ciudad. 2024. Immediate Impact of an 8-Week Virtual Reality Educational Program on Burnout and Work Engagement Among Health Care Professionals: Pre-Post Pilot Study. *JMIR XR and Spatial Computing* 1: e55678. https://doi.org/10.2196/55678

[45] Jose Ferrer Costa, Nuria Moran, Carlos Garcia Marti, Leomar Javier Colmenares Hernandez, Florin Radu Ciorba Ciorba, and Maria Jose Ciudad. 2024. Immediate Impact of an 8-Week Virtual Reality Educational Program on Burnout and Work Engagement Among Health Care Professionals: Pre-Post Pilot Study. *JMIR XR and Spatial Computing* 1: e55678. https://doi.org/10.2196/55678

[46] Francesco Fischetti, Ilaria Pepe, Gianpiero Greco, Maurizio Ranieri, Luca Poli, Stefania Cataldi, and Luigi Vimercati. 2024. Ten-Minute Physical Activity Breaks Improve Attention and Executive Functions in Healthcare Workers. *Journal of Functional Morphology and Kinesiology* 9, 2: 102. https://doi.org/10.3390/jfmk9020102

[47] Theresa Fleming, Lynda Bavin, Mathijs Lucassen, Karolina Stasiak, Sarah Hopkins, and Sally Merry. 2018. Beyond the Trial: Systematic Review of Real-World Uptake and Engagement With Digital Self-Help Interventions for Depression, Low Mood, or Anxiety. *Journal of Medical Internet Research* 20, 6: e199. https://doi.org/10.2196/jmir.9275

[48] Zhongfang Fu, Huibert Burger, Retha Arjadi, and Claudi L H Bockting. 2020. Effectiveness of digital psychological interventions for mental health problems in low-income and middle-income countries: a systematic review and meta-analysis. *The Lancet Psychiatry* 7, 10: 851–864. https://doi.org/10.1016/S2215-0366(20)30256-X

[49] Patricia L. Gerbarg, Felicity Dickson, Vincent A. Conte, and Richard P. Brown. 2023. Breath-centered virtual mind-body medicine reduces COVID-related stress in women healthcare workers of the Regional Integrated Support for Education in Northern Ireland: a single group study. *Frontiers in Psychiatry* 14: 1199819. https://doi.org/10.3389/fpsyt.2023.1199819

[50] Nishmi Gunasingam, Kharis Burns, James Edwards, Michael Dinh, and Merrilyn Walton. 2015. Reducing stress and burnout in junior doctors: the impact of debriefing sessions. *Postgraduate Medical Journal* 91, 1074: 182–187. https://doi.org/10.1136/postgradmedj-2014-132847

[51] Luke Haliburton, Benedikt Pirker, Paolo Holinski, Albrecht Schmidt, Pawel W. Wozniak, and Matthias Hoppe. 2023. VR-Hiking: Physical Exertion Benefits Mindfulness and Positive Emotions in Virtual Reality. *Proceedings of the ACM on Human-Computer Interaction* 7, MHCI: 1–17. https://doi.org/10.1145/3604263

[52] Louise H. Hall, Judith Johnson, Ian Watt, Anastasia Tsipa, and Daryl B. O'Connor. 2016. Healthcare Staff Wellbeing, Burnout, and Patient Safety: A Systematic Review. *PLOS ONE* 11, 7: e0159015. https://doi.org/10.1371/journal.pone.0159015

[53] Ping-Hsuan Han, Yang-Sheng Chen, Yilun Zhong, Han-Lei Wang, and Yi-Ping Hung. 2017. My Tai-Chi coaches: an augmented-learning tool for practicing Tai-Chi Chuan. In *Proceedings of the 8th Augmented Human International Conference*, 1–4. https://doi.org/10.1145/3041164.3041194

[54] Yuzhu Hao, Qiuxia Wu, Xiaoyang Luo, Shubao Chen, Chang Qi, Jiang Long, Yifan Xiong, Yanhui Liao, and Tieqiao Liu. 2020. Mental Health Literacy of Non-mental Health Nurses: A Mental Health Survey in Four General Hospitals in Hunan Province, China. *Frontiers in Psychiatry* 11: 507969. https://doi.org/10.3389/fpsyt.2020.507969

[55] Marja Harjumaa, Pilvikki Absetz, Miikka Ermes, Elina Mattila, Reija Männikkö, Tanja Tilles-Tirkkonen, Niina Lintu, Ursula Schwab, Adil Umer, Juha Leppänen, and Jussi Pihlajamäki. 2020. Internet-Based Lifestyle Intervention to Prevent Type 2 Diabetes Through Healthy Habits: Design and 6-Month Usage Results of Randomized Controlled Trial. *JMIR Diabetes* 5, 3: e15219. https://doi.org/10.2196/15219

[56] Rebekah K. Hersch, Royer F. Cook, Diane K. Deitz, Seth Kaplan, Daniel Hughes, Mary Ann Friesen, and Maria Vezina. 2016. Reducing nurses' stress: A randomized controlled trial of a web-based stress management program for nurses. *Applied Nursing Research* 32: 18–25. https://doi.org/10.1016/j.apnr.2016.04.003

[57] Esther Howe, Jina Suh, Mehrab Bin Morshed, Daniel McDuff, Kael Rowan, Javier Hernandez, Marah Ihab Abdin, Gonzalo Ramos, Tracy Tran, and Mary P Czerwinski. 2022. Design of Digital Workplace Stress-Reduction Intervention Systems: Effects of Intervention Type and Timing. In *CHI Conference on Human Factors in Computing Systems*, 1–16. https://doi.org/10.1145/3491102.3502027

[58] Jun Huang, Qianwen Tan, and Qi Fang. 2024. VitalStep: Revitalizing Elderly Health and Connectivity with E-Square-Dance. In *Extended Abstracts of the CHI Conference on Human Factors in Computing Systems*, 1–7. https://doi.org/10.1145/3613905.3647978

[59] Kiran Ijaz, Naseem Ahmadpour, Yifan Wang, and Rafael A. Calvo. 2020. Player Experience of Needs Satisfaction (PENS) in an Immersive Virtual Reality Exercise Platform Describes Motivation and Enjoyment. *International Journal of Human–Computer Interaction* 36, 13: 1195–1204. https://doi.org/10.1080/10447318.2020.1726107

[60] Jacinta Jardine, Camille Nadal, Sarah Robinson, Angel Enrique, Marcus Hanratty, and Gavin Doherty. 2023. Between Rhetoric and Reality: Real-world Barriers to Uptake and Early Engagement in Digital Mental Health Interventions. *ACM Transactions on Computer-Human Interaction*: 3635472. https://doi.org/10.1145/3635472

[61] Rebecca M. Jedwab, Alison M. Hutchinson, Elizabeth Manias, Rafael A. Calvo, Naomi Dobroff, Nicholas Glozier, and Bernice Redley. 2021. Nurse Motivation, Engagement and Well-Being before an Electronic Medical Record System Implementation: A Mixed Methods Study. *International Journal of Environmental Research and Public Health* 18, 5: 2726. https://doi.org/10.3390/ijerph18052726

[62] Hye-Young Jo, Laurenz Seidel, Michel Pahud, Mike Sinclair, and Andrea Bianchi. 2023. FlowAR: How Different Augmented Reality Visualizations of Online Fitness Videos Support Flow for At-Home Yoga Exercises. In *Proceedings of the 2023 CHI Conference on Human Factors in Computing Systems*, 1–17. https://doi.org/10.1145/3544548.3580897

[63] Su-Eun Jung, Da-Jung Ha, Jung-Hyun Park, Boram Lee, Myo-Sung Kim, Kyo-Lin Sim, Yung-Hyun Choi, and Chan-Young Kwon. 2021. The Effectiveness and Safety of Mind-Body Modalities for Mental Health of Nurses in Hospital Setting: A Systematic Review. *International Journal of Environmental Research and Public Health* 18, 16: 8855. https://doi.org/10.3390/ijerph18168855

[64] Anne C. Krendl and Bernice A. Pescosolido. 2020. Countries and Cultural Differences in the Stigma of Mental Illness: The East–West Divide. *Journal of Cross-Cultural Psychology* 51, 2: 149–167. https://doi.org/10.1177/0022022119901297

[65] Sarah Angela Kriakous, Katie Ann Elliott, Carolien Lamers, and Robin Owen. 2021. The Effectiveness of Mindfulness-Based Stress Reduction on the Psychological Functioning of Healthcare Professionals: a Systematic Review. *Mindfulness* 12, 1: 1–28. https://doi.org/10.1007/s12671-020-01500-9

[66] Sonia Lippke, Lingling Gao, Franziska Maria Keller, Petra Becker, and Alina Dahmen. 2021. Adherence With Online Therapy vs Face-to-Face Therapy and With Online Therapy vs Care as Usual: Secondary Analysis of Two Randomized Controlled Trials. *Journal of Medical Internet Research* 23, 11: e31274. https://doi.org/10.2196/31274

[67] Yolanda López-Del-Hoyo, Selene Fernández-Martínez, Adrián Pérez-Aranda, Alberto Barceló-Soler, Marco Bani, Selena Russo, Fernando Urcola-Pardo, Maria Grazia Strepparava, and Javier García-Campayo. 2023. Effects of eHealth interventions on stress reduction and mental health promotion in healthcare professionals: A systematic review. *Journal of Clinical Nursing* 32, 17–18: 5514–5533. https://doi.org/10.1111/jocn.16634

[68] Michelle Luken and Amanda Sammons. 2016. Systematic Review of Mindfulness Practice for Reducing Job Burnout. *The American Journal of Occupational Therapy* 70, 2: 7002250020p1-7002250020p10. https://doi.org/10.5014/ajot.2016.016956

[69] Lumivero. Nvivo. Retrieved from https://lumivero.com/products/nvivo/

[70] Marie Michele Macaron, Omotayo Ayomide Segun-Omosehin, Reem H. Matar, Azizullah Beran, Hayato Nakanishi, Christian A. Than, and Osama A. Abulseoud. 2023. A systematic review and meta analysis on burnout in physicians during the COVID-19 pandemic: A hidden healthcare crisis. *Frontiers in Psychiatry* 13: 1071397. https://doi.org/10.3389/fpsyt.2022.1071397

[71] Preetham Madapura Nagaraj, Wen Mo, and Catherine Holloway. 2024. Mindfulness-based Embodied Tangible Interactions for Stroke Rehabilitation at Home. In *Proceedings of the CHI Conference on Human Factors in Computing Systems*, 1–16. https://doi.org/10.1145/3613904.3642463

[72] Juan F. Maestre, Daria V. Groves, Megan Furness, and Patrick C. Shih. 2023. "It's like With the Pregnancy Tests": Co-design of Speculative Technology for Public HIV-related Stigma and its Implications for Social Media. In *Proceedings of the 2023 CHI Conference on Human Factors in Computing Systems*, 1–21. https://doi.org/10.1145/3544548.3581033

[73] Raju Maharjan, Per Bækgaard, and Jakob E. Bardram. 2019. "Hear me out": smart speaker based conversational agent to monitor symptoms in mental health. In *Adjunct Proceedings of the 2019 ACM International Joint Conference on Pervasive and Ubiquitous Computing and Proceedings of the 2019 ACM International Symposium on Wearable Computers*, 929–933. https://doi.org/10.1145/3341162.3346270

[74] Giuseppa Maresca, Francesco Corallo, Giulia Catanese, Caterina Formica, and Viviana Lo Buono. 2022. Coping Strategies of Healthcare Professionals with Burnout Syndrome: A Systematic Review. *Medicina* 58, 2: 327. https://doi.org/10.3390/medicina58020327

[75] David Marshall, Grainne Donohue, Jean Morrissey, and Brendan Power. 2018. Evaluation of a Tai Chi Intervention to Promote Well-Being in Healthcare Staff: A Pilot Study. *Religions* 9, 2: 35. https://doi.org/10.3390/rel9020035

[76] Ines Miguel-Alonso, Bruno Rodriguez-Garcia, David Checa, and Andres Bustillo. 2023. Countering the Novelty Effect: A Tutorial for Immersive Virtual Reality Learning Environments. *Applied Sciences* 13, 1: 593. https://doi.org/10.3390/app13010593

[77] David C. Mohr, Ken R. Weingardt, Madhu Reddy, and Stephen M. Schueller. 2017. Three Problems With Current Digital Mental Health Research . . . and Three Things We Can Do About Them. *Psychiatric Services* 68, 5: 427–429. https://doi.org/10.1176/appi.ps.201600541

[78] Jesus Montero-Marin, Jorge Gaete, Ricardo Araya, Marcelo Demarzo, Rick Manzanera, Melchor Álvarez De Mon, and Javier García-Campayo. 2018. Impact of a Blended Web-Based Mindfulness Programme for General Practitioners: a Pilot Study. *Mindfulness* 9, 1: 129–139. https://doi.org/10.1007/s12671-017-0752-8

[79] Gaye Moore, Helen Wilding, Kathleen Gray, and David Castle. 2019. Participatory Methods to Engage Health Service Users in the Development of Electronic Health Resources: Systematic Review. *Journal of Participatory Medicine* 11, 1: e11474. https://doi.org/10.2196/11474

[80] Arun Nagargoje, Karl Maybach, and Tomas Sokoler. 2012. Social yoga mats: designing for exercising/socializing synergy. In *Proceedings of the Sixth International Conference on Tangible, Embedded and Embodied Interaction*, 87–90. https://doi.org/10.1145/2148131.2148151

[81] Matthew Naylor, Ben Morrison, Brad Ridout, and Andrew Campbell. 2019. Augmented Experiences: Investigating the Feasibility of Virtual Reality as Part of a Workplace Wellbeing Intervention. *Interacting with Computers* 31, 5: 507–523. https://doi.org/10.1093/iwc/iwz033

[82] Francisco Nunes, Nervo Verdezoto, Geraldine Fitzpatrick, Morten Kyng, Erik Grönvall, and Cristiano Storni. 2015. Self-Care Technologies in HCI: Trends, Tensions, and Opportunities. *ACM Transactions on Computer-Human Interaction* 22, 6: 1–45. https://doi.org/10.1145/2803173

[83] Yoshihiro Okada, Takayuki Ogata, and Hiroyuki Matsuguma. 2016. Component-Based Approach for Prototyping of Tai Chi-Based Physical Therapy Game and Its Performance Evaluations. *Computers in Entertainment* 14, 1: 1–20. https://doi.org/10.1145/2735383

[84] Maria Panagioti, Efharis Panagopoulou, Peter Bower, George Lewith, Evangelos Kontopantelis, Carolyn Chew-Graham, Shoba Dawson, Harm van Marwijk, Keith Geraghty, and Aneez Esmail. 2017. Controlled Interventions to Reduce Burnout in Physicians: A Systematic Review and Meta-analysis. *JAMA Internal Medicine* 177, 2: 195. https://doi.org/10.1001/jamainternmed.2016.7674

[85] SoHyun Park, Anja Thieme, Jeongyun Han, Sungwoo Lee, Wonjong Rhee, and Bongwon Suh. 2021. "I wrote as if I were telling a story to someone I knew.": Designing Chatbot Interactions for Expressive Writing in Mental Health. In *Designing Interactive Systems Conference 2021*, 926–941. https://doi.org/10.1145/3461778.3462143

[86] King Pascual, Amiad Fredman, Athanasios Naum, Chaitrali Patil, and Neal Sikka. 2023. Should Mindfulness for Health Care Workers Go Virtual? A Mindfulness-Based Intervention Using Virtual Reality and Heart Rate Variability in the Emergency Department. *Workplace Health & Safety* 71, 4: 188–194. https://doi.org/10.1177/21650799221123258

[87] Sachin R. Pendse, Kate Niederhoffer, and Amit Sharma. 2019. Cross-Cultural Differences in the Use of Online Mental Health Support Forums. *Proceedings of the ACM on Human-Computer Interaction* 3, CSCW: 1–29. https://doi.org/10.1145/3359169

[88] Sachin R Pendse, Daniel Nkemelu, Nicola J Bidwell, Sushrut Jadhav, Soumitra Pathare, Munmun De Choudhury, and Neha Kumar. 2022. From Treatment to Healing:Envisioning a Decolonial Digital Mental Health. In *CHI Conference on Human Factors in Computing Systems*, 1–23. https://doi.org/10.1145/3491102.3501982

[89] Giselle K. Perez, Vivian Haime, Vicki Jackson, Eva Chittenden, Darshan H. Mehta, and Elyse R. Park. 2015. Promoting Resiliency among Palliative Care Clinicians: Stressors, Coping Strategies, and Training Needs. *Journal of Palliative Medicine* 18, 4: 332–337. https://doi.org/10.1089/jpm.2014.0221

[90] Dorian Peters. 2023. Wellbeing Supportive Design – Research-Based Guidelines for Supporting Psychological Wellbeing in User Experience. *International Journal of Human–Computer Interaction* 39, 14: 2965–2977. https://doi.org/10.1080/10447318.2022.2089812

[91] Dorian Peters and Rafael A. Calvo. 2023. Self-Determination Theory and Technology Design. In *The Oxford Handbook of Self-Determination Theory* (1st ed.), Richard M. Ryan (ed.). Oxford University Press, 978–999. https://doi.org/10.1093/oxfordhb/9780197600047.013.49

[92] Dorian Peters, Rafael A. Calvo, and Richard M. Ryan. 2018. Designing for Motivation, Engagement and Wellbeing in Digital Experience. *Frontiers in Psychology* 9: 797. https://doi.org/10.3389/fpsyg.2018.00797

[93] Dorian Peters, Sharon Davis, Rafael Alejandro Calvo, Susan M Sawyer, Lorraine Smith, and Juliet M Foster. 2017. Young People's Preferences for an Asthma Self-Management App Highlight Psychological Needs: A Participatory Study. *Journal of Medical Internet Research* 19, 4: e113. https://doi.org/10.2196/jmir.6994

[94] Claudette Pretorius, Darragh McCashin, Naoise Kavanagh, and David Coyle. 2020. Searching for Mental Health: A Mixed-Methods Study of Young People's Online Help-seeking. In *Proceedings of the 2020 CHI Conference on Human Factors in Computing Systems*, 1–13. https://doi.org/10.1145/3313831.3376328

[95] Jochen Profit, Kathryn C. Adair, Xin Cui, Briana Mitchell, Debra Brandon, Daniel S. Tawfik, Joseph Rigdon, Jeffrey B. Gould, Henry C. Lee, Wendy L. Timpson, Martin J. McCaffrey, Alexis S. Davis, Mohan Pammi, Melissa Matthews, Ann R. Stark, Lu-Ann Papile, Eric Thomas, Michael Cotten, Amir Khan, and J. Bryan Sexton. 2021. Randomized controlled trial of the "WISER" intervention to reduce healthcare worker burnout. *Journal of Perinatology* 41, 9: 2225–2234. https://doi.org/10.1038/s41372-021-01100-y

[96] Ranxing Technology. WJX. Retrieved from https://www.wjx.cn/

[97] Michael C Robertson, Elizabeth J Lyons, Yue Liao, Miranda L Baum, and Karen M Basen-Engquist. 2020. Gamified Text Messaging Contingent on Device-Measured Steps: Randomized Feasibility Study of a Physical Activity Intervention for Cancer Survivors. *JMIR mHealth and uHealth* 8, 11: e18364. https://doi.org/10.2196/18364

[98] Maria Francesca Rossi, Maria Rosaria Gualano, Nicola Magnavita, Umberto Moscato, Paolo Emilio Santoro, and Ivan Borrelli. 2023. Coping with burnout and the impact of the COVID-19 pandemic on workers' mental health: A systematic review. *Frontiers in Psychiatry* 14: 1139260. https://doi.org/10.3389/fpsyt.2023.1139260

DIS '25, July 05–09, 2025, Funchal, Portugal

Zheyuan Zhang et al.

[99] Lisa S. Rotenstein, Matthew Torre, Marco A. Ramos, Rachael C. Rosales, Constance Guille, Srijan Sen, and Douglas A. Mata. 2018. Prevalence of Burnout Among Physicians: A Systematic Review. *JAMA* 320, 11: 1131. https://doi.org/10.1001/jama.2018.12777

[100] Richard M. Ryan (ed.). 2023. *The Oxford Handbook of Self-Determination Theory*. Oxford University Press. https://doi.org/10.1093/oxfordhb/9780197600047.001.0001

[101] Richard M Ryan and Edward L Deci. 2000. Self-Determination Theory and the Facilitation of Intrinsic Motivation, Social Development, and Well-Being. *American Psychologist*: 11.

[102] Richard M. Ryan and Edward L. Deci. 2020. Intrinsic and extrinsic motivation from a self-determination theory perspective: Definitions, theory, practices, and future directions. *Contemporary Educational Psychology* 61: 101860. https://doi.org/10.1016/j.cedpsych.2020.101860

[103] Richard M. Ryan, C. Scott Rigby, and Andrew Przybylski. 2006. The Motivational Pull of Video Games: A Self-Determination Theory Approach. *Motivation and Emotion* 30, 4: 344–360. https://doi.org/10.1007/s11031-006-9051-8

[104] Cassandra E.L. Seah, Sijin Sun, Zheyuan Zhang, Talya Porat, Andrew Waterhouse, and Rafael A. Calvo. 2022. Using a User Centered Design Approach to Design Mindfulness Conversational Agent for Persons with Dementia and their Caregivers. In *Proceedings of the 2022 ACM International Joint Conference on Pervasive and Ubiquitous Computing*, 207–210. https://doi.org/10.1145/3544793.3563398

[105] Matthew Seita, Sooyeon Lee, Sarah Andrew, Kristen Shinohara, and Matt Huenerfauth. 2022. Remotely Co-Designing Features for Communication Applications using Automatic Captioning with Deaf and Hearing Pairs. In *CHI Conference on Human Factors in Computing Systems*, 1–13. https://doi.org/10.1145/3491102.3501843

[106] Mehrdad Sharifi, Ali Akbar Asadi-Pooya, and Razieh Sadat Mousavi-Roknabadi. 2020. Burnout among Healthcare Providers of COVID-19; a Systematic Review of Epidemiology and Recommendations: Burnout in healthcare providers. *Archives of Academic Emergency Medicine* 9, 1: e7. https://doi.org/10.22037/aaem.v9i1.1004

[107] Jennifer L. Shaw-Metz. 2023. Coming up for air: Breathwork practice for stress management in the healthcare setting. *Journal of Interprofessional Education & Practice* 30: 100594. https://doi.org/10.1016/j.xjep.2022.100594

[108] Rui She, Xiaohui Wang, Zhoubin Zhang, Jinghua Li, Jingdong Xu, Hua You, Yan Li, Yuan Liang, Shan Li, Lina Ma, Xinran Wang, Xiuyuan Chen, Peien Zhou, Joseph Lau, Yuantao Hao, Huan Zhou, and Jing Gu. 2021. Mental Health Help-Seeking and Associated Factors Among Public Health Workers During the COVID-19 Outbreak in China. *Frontiers in Public Health* 9: 622677. https://doi.org/10.3389/fpubh.2021.622677

[109] Grace Shin, Yuanyuan Feng, Mohammad Hossein Jarrahi, and Nicci Gafinowitz. 2019. Beyond novelty effect: a mixed-methods exploration into the motivation for long-term activity tracker use. *JAMIA Open* 2, 1: 62–72. https://doi.org/10.1093/jamiaopen/ooy048

[110] Sang-Wha Sien, Jessica Y. Ahn, and Joanna McGrenere. 2023. Co-designing Mental Health Technologies with International University Students in Canada. *Proceedings of the ACM on Human-Computer Interaction* 7, CSCW2: 1–25. https://doi.org/10.1145/3610049

[111] Elizabeth Stratton, Amit Lampit, Isabella Choi, Hanna Malmberg Gavelin, Melissa Aji, Jennifer Taylor, Rafael A Calvo, Samuel B Harvey, and Nick Glozier. 2022. Trends in Effectiveness of Organizational eHealth Interventions in Addressing Employee Mental Health: Systematic Review and Meta-analysis. *Journal of Medical Internet Research* 24, 9: e37776. https://doi.org/10.2196/37776

[112] Sijin Sun, Zheyuan Zhang, Mu Tian, Celine Mougenot, Nick Glozier, and Rafael A Calvo. 2022. Preferences for a Mental Health Support Technology Among Chinese Employees: Mixed Methods Approach. *JMIR Human Factors* 9, 4: e40933. https://doi.org/10.2196/40933

[113] Tencent. WeChat. Retrieved from https://www.wechat.com/

[114] Lia Tirabeni. 2023. Bounded Well-Being: Designing Technologies for Workers' Well-Being in Corporate Programmes. *Work, Employment and Society*: 09500170231203113. https://doi.org/10.1177/09500170231203113

[115] Edwin N. Torres and Tingting Zhang. 2021. The impact of wearable devices on employee wellness programs: A study of hotel industry workers. *International Journal of Hospitality Management* 93: 102769. https://doi.org/10.1016/j.ijhm.2020.102769

[116] Evlalia Touloudi, Mary Hassandra, Evangelos Galanis, Marios Goudas, and Yannis Theodorakis. 2022. Applicability of an Immersive Virtual Reality Exercise Training System for Office Workers during Working Hours. *Sports* 10, 7: 104. https://doi.org/10.3390/sports10070104

[117] Laia Turmo Vidal, Elena Márquez Segura, Christopher Boyer, and Annika Waern. 2019. Enlightened Yoga: Designing an Augmented Class with Wearable Lights to Support Instruction. In *Proceedings of the 2019 on Designing Interactive Systems Conference*, 1017–1031. https://doi.org/10.1145/3322276.3322338

[118] April Tyack and Elisa D. Mekler. 2020. Self-Determination Theory in HCI Games Research: Current Uses and Open Questions. In *Proceedings of the 2020 CHI Conference on Human Factors in Computing Systems*, 1–22. https://doi.org/10.1145/3313831.3376723

[119] Timothy J. Usset, R. Greg Stratton, Sarah Knapp, Gabrielle Schwartzman, Sunil K. Yadav, Benjamin J. Schaefer, J. Irene Harris, and George Fitchett. 2024. Factors Associated With Healthcare Clinician Stress and Resilience: A Scoping Review. *Journal of Healthcare Management* 69, 1: 12–28. https://doi.org/10.1097/JHM-D-23-00020

[120] Madhan Kumar Vasudevan, Shu Zhong, Jan Kučera, Desiree Cho, and Marianna Obrist. 2023. MindTouch: Effect of Mindfulness Meditation on Mid-Air Tactile Perception. In *Proceedings of the 2023 CHI Conference on Human Factors in Computing Systems*, 1–12. https://doi.org/10.1145/3544548.3581238

[121] Lin Wang and Weifeng Deng. 2023. Research on the Auxiliary Training System of Tai Chi Fitness Qigong Based on Computer 3D Image Vision Technology. In *2023 IEEE International Conference on Image Processing and Computer Applications (ICIPCA)*, 177–182. https://doi.org/10.1109/ICIPCA59209.2023.10257873

[122] Colin P West, Liselotte N Dyrbye, Patricia J Erwin, and Tait D Shanafelt. 2016. Interventions to prevent and reduce physician burnout: a systematic review and meta-analysis. *The Lancet* 388, 10057: 2272–2281. https://doi.org/10.1016/S0140-6736(16)31279-X

[123] Rachel Woo, Daniel Harley, and James R Wallace. 2024. "I'm not alone in that battle": Designing Mobile AR for Mental Health Communication and Community Connectedness. In *Designing Interactive Systems Conference*, 328–342. https://doi.org/10.1145/3643834.3661626

[124] Chenghui Yang, Bo Zhou, Jinyu Wang, and Shuya Pan. 2021. The effect of a short-term Balint group on the communication ability and self-efficacy of pre-examination and triage nurses during COVID-19. *Journal of Clinical Nursing* 30, 1–2: 93–100. https://doi.org/10.1111/jocn.15489

[125] Lucy Yardley, Leanne Morrison, Katherine Bradbury, and Ingrid Muller. 2015. The Person-Based Approach to Intervention Development: Application to Digital Health-Related Behavior Change Interventions. *Journal of Medical Internet Research* 17, 1: e30. https://doi.org/10.2196/jmir.4055

[126] Hui Ye, Jiaye Leng, Pengfei Xu, Karan Singh, and Hongbo Fu. 2024. ProInterAR: A Visual Programming Platform for Creating Immersive AR Interactions. In *Proceedings of the CHI Conference on Human Factors in Computing Systems*, 1–15. https://doi.org/10.1145/3613904.3642527

[127] N.L. Yeo, M.P. White, I. Alcock, R. Garside, S.G. Dean, A.J. Smalley, and B. Gatersleben. 2020. What is the best way of delivering virtual nature for improving mood? An experimental comparison of high definition TV, 360° video, and computer generated virtual reality. *Journal of Environmental Psychology* 72: 101500. https://doi.org/10.1016/j.jenvp.2020.101500

[128] Bin Yu, Jun Hu, Mathias Funk, and Loe Feijs. 2018. DeLight: biofeedback through ambient light for stress intervention and relaxation assistance. *Personal and Ubiquitous Computing* 22, 4: 787–805. https://doi.org/10.1007/s00779-018-1141-6

[129] Zheyuan Zhang, Sijin Sun, Laura Moradbakhti, Andrew Hall, Celine Mougenot, Juan Chen, and Rafael A Calvo. 2025. Health Care Professionals' Engagement With Digital Mental Health Interventions in the United Kingdom and China: Mixed Methods Study on Engagement Factors and Design Implications. *JMIR Mental Health* 12: e67190–e67190. https://doi.org/10.2196/67190

# Appendices
# Appendix A

**Table 4: Co-design Workshop Plan**

| Phase | Activities | Tools and materials |
|---|---|---|
| Phase 1 | General introduction to the workshop aims and process (3 mins) | Presentation slides |
| | Brief Reflection (3 min): As a healthcare professional, how do you cope with stress and burnout? Is there any product, digital or non-digital, that you are using to support relaxation? | N/A |
| | Introduction to existing evidence-based burnout and wellbeing interventions for HCPs, followed by Q&A sessions (3 mins) | Presentation slides and printed design references |
| | Introduction to current technologies available for wellbeing interventions, followed by Q&A sessions (3 mins) | Presentation slides and printed design references |
| | Idea generation on technology-based interventions (10 mins) | Printed design templates, white board, pens and markers |
| | Presentation and group discussion (15 mins individual presentation) | |
| | Idea evaluation and voting (5 mins) | Voting stickers |
| Phase 2 (8 weeks after Phase 1) | Opening introduction to the workshop aim and the prototype (5 min) | Presentation slides |
| | Prototype experience sessions, with participants divided into small groups of 2-4 people, each session took around 2 minutes (10 mins) | Qigong exercise interactive prototype (Screen, PC, and Microsoft Kinect Sensor) |
| | Reflection and discussion on the positive and negative experiences (10 mins) | Pens and papers |
| | Introduction to the three design templates (5 mins) | Presentation slides |
| | Design iteration / redesign session (20 mins) | Printed design templates, markers and pens |
| | Idea presentation and discussion (40 mins) | N / A |

# Appendix B

**Table 5: Ideas generated by participants in Phase 1**

| Participant ID | Generated Ideas |
| --- | --- |
| P1 | 1. Yoga Classroom: Use projection mapping to transform the current staff room into a yoga classroom<br>2. VR Mindfulness: Relax and do mindfulness meditation wearing a VR headset showing nature scenes, like seaside, jungle, and galaxy etc. |
| P2 | 3. 5D Experience: Use projection or displays, along with other tools to stimulate multiple senses and create the "5D experience", where you experience a film or environment in extra reality. |
| P3 | 4. Relaxation TV: A big TV in the staff room with content on yoga, mindfulness movements or intensive exercises to support relaxation. Motion sensors can be connected to detect multiple players movement. This can help de-stress and create moment to togetherness among colleagues. |
| P4 | 5. On-screen Yoga Coach: A yoga learning software on TV, with motion sensor that detect learners' body movement.<br>6. Versatile Staff Room: Use immersive projections to turn the staff room into a versatile and shifting environment, to match different wellbeing needs. |
| P5 | 7. VR Venture: Use VR to help relaxation by helping HCPs get out of the clinical environment, forget they are HCPs for a while and even engage with extreme sports in the virtual environment.<br>8. VR-based LRP (live role-play game): just like those LRP experience in the physical rooms, P5 liked to engage with LRP games even for a short while during breaks, for quick relaxation, distraction from stressors and the appealing stories. |
| P6 | 9. Magic Mats: An interactive mat that detect your foot pressure and body weight, and responds with on-screen graphics, light and sound. This mat can facilitate active exercise within the workplace without occupying a large space.<br>10. AR/VR Art Therapy: experience art therapy using AR/VR, in which you can observe an art piece being made, or create your own work, with music and sound effect to form relaxing experiences. |
| P7 | 11. VR Exercise and Tai-chi: Create a VR intervention to help practice Tai-chi, in order to relax your body, especially shoulder and neck, as P7 always got neck and shoulder pain due to work. |
| P8 | 12. Relaxation Room with Massage Chairs: Refurbish the current staff room into a relaxation space with soundscape and smart lighting, as long as massage chairs to help relax physically, especially shoulder and neck.<br>13. Karaoke Pod: Sound insulated Karaoke pod, where staff can go singing without disturbing others. |
| P9 | 14. Immersive Art Therapy: Create an immersive space for art and music therapy, to support short sessions focusing on art experiences as a quick de-stress mechanism.<br>15. Virtual Boardgame: Use AR to create board games, or LRP games, that can be played among several staff, to increase communication among staff.<br>16. Motion Sensing Game: Use motion-sensing technologies, to create a quick physical game where you need to reach goals and get rewards. This facilitates staff physical exercise and motivate the next engagement. |
| P10 | 17. Digital message board: A message board on an iPad, or on a screen, where staff can leave messages and communicate anonymously to encourage a supportive team spirit. |
| P11 | 18. Immersive Room: Upgrade the staff room to a 360° immersive room, where staff can relax, watch TV, and immerse in different realistic scenes.<br>19. Mini Stories in Cardboard VR: Use carboard VR to watch mini-series, as those popular on Douyin and Kuaishou (both short-video platforms), as many staff liked to binge watch short stories to relax. |
| P12 | 20. Immersive Nature: Use VR, 360° displays to create an immersive garden for brief mindfulness and relaxation.<br>21. Dance Machine: Similar to the interactive dancing mats or pads in the arcade, a small machine that can facilitate playful movement with sound and light feedback, and scores, rankings etc. |
| P13 | 22. VR Tai-chi and Qigong: Doing Qigong and Tai-chi in a virtual environment, use VR controllers and body recognition to practice key movements. P13 mentioned that such movements can benefit staff who have just finished their night shift.<br>23. WeChat Therapy Chatbot: A chatbot built for dialectical therapy, which you can turn to for a chat and vent out, on WeChat. |
| P14 | 24. Sand Painting: Use technology like AR, VR or touch screen to allow HCPs to do sand painting. Sand painting, according to P14, was quite relaxing and mesmerizing.<br>25. Breathwork Robot: A smart speaker, or a robot therapist which help and guide staff on breathwork trainings, in order to learn efficient skills to relax. |
| P15 | 26. Basketball Game: A virtual basketball game with tangible tools to mimic the touch and bounce of a real basketball. This allows HCPs to exercise and energize without taking too much space. |
| | 27. Motion Sensing Mini Sports Game: Use motion sensing to create interactive sports game in little "bites", where staff can quickly engage and exercise. |
| | 28. Staff Room Cinema: Use projection or surrounding displays to set up a cinema for staff to relax and unwind. |
| P16 | 29. Tai-chi Game: Motion-sensing game that support practice of Tai-chi, as P16 always wanted to join a Tai-chi skill class but was too busy to attend. |
| | 30. Boxing Sandbag: an interactive sandbag in the staff room that help staff to practice boxing or to vent negative emotions. |